\documentclass{emulateapj}

\slugcomment{To Appear in the Astrophysical Journal}
\shorttitle{RMHD simulations of inflow and outflow around BH}
\shortauthors{OHSUGA \& MINESHIGE}

\newcommand{\lsim}{\raisebox{0.3mm}{\em $\, <$} \hspace{-3.3mm}
\raisebox{-1.8mm}{\em $\sim \,$}}
\newcommand{\gsim}{\raisebox{0.3mm}{\em $\, >$} \hspace{-3.3mm}
\raisebox{-1.8mm}{\em $\sim \,$}}
\newcommand{\bma}[1]{\mbox{\boldmath $#1$}}

\begin{document}

\title
{GLOBAL STRUCTURE OF THREE DISTINCT ACCRETION
FLOWS AND OUTFLOWS AROUND BLACK HOLES
THROUGH TWO-DIMENSIONAL
RADIATION-MAGNETOHYDRODYNAMIC
SIMULATIONSGlobal
}

\author
{Ken {\sc Ohsuga}\altaffilmark{1, 2}
{\sc and} 
 Shin {\sc Mineshige}\altaffilmark{3}
}

\affil{\altaffilmark{1} National Astronomical Observatory of Japan, Osawa, Mitaka, Tokyo 181-8588, Japan}
\affil{\altaffilmark{2} School of Physical Sciences,Graduate University of
Advanced Study (SOKENDAI), Shonan Village, Hayama, Kanagawa 240-0193, Japan}
\affil{\altaffilmark{3} Department of Astronomy, Graduate School of Science, Kyoto University, Kyoto 606-8502, Japan}



\begin{abstract}
We present the detailed global structure of black hole accretion 
flows and outflows through newly performed two-dimensional
radiation-magnetohydrodynamic simulations.
By starting from a torus threaded with weak toroidal magnetic fields
and by controlling the central density of the initial torus, $\rho_0$,
we can reproduce three distinct modes of accretion flow.
In model A with the highest central density,
an optically and geometrically thick supercritical accretion disk is
created.
The radiation force greatly exceeds the gravity above the disk surface,
thereby driving a strong outflow (or jet).
Because of the mild beaming,
the apparent (isotropic) photon luminosity is $\sim 22L_{\rm E}$
(where $L_{\rm E}$ is the Eddington luminosity) in the face-on view.
Even higher apparent luminosity is feasible if we increase the flow density.
In model B with a moderate density,
radiative cooling of the accretion flow is so efficient
that a standard-type, cold, and geometrically thin disk is formed
at radii greater than $\sim 7R_{\rm S}$
(where $R_{\rm S}$ is the Schwarzschild radius), while
the flow is radiatively inefficient otherwise.
The magnetic-pressure-driven disk wind appears in this model.
In model C the density is too low for the flow to be radiatively efficient.
The flow thus becomes radiatively inefficient accretion flow,
which is geometrically thick and optically thin.
The magnetic-pressure force, in cooperation with the gas-pressure force,
drives outflows from the disk surface, and
the flow releases its energy via jets rather than via radiation.
Observational implications are briefly discussed.
\end{abstract}


\section{Introduction}

Black hole accretion disks provide the most powerful 
energy-production mechanism in the universe.
However, theoretical development in this area is rather behind, 
compared with that of the stars.
The central engine of the disks was identified 
as having magnetic origin \citep{BH91} in the late 1990s after 
extensive investigations had been carried out by many authors.
Yet, no complete picture of magnetized accretion flow and 
outflow has yet been obtained at this moment. 
We may thus say that
the theory of the accretion disks is now in a similar situation 
to that of the stars in the 1940s$-$1950s.


Before the identification of the central engine, 
one-dimensional accretion disk models were constructed
based on the phenomenological $\alpha$-viscosity prescription, 
whereby the viscous torque is proportional to the pressure,
after the pioneering work by \citep{SS73}.
The standard disk model was first established as a model for
accretion disks at moderately high luminosities,
followed by various disk models, including the slim disk model 
and the radiatively inefficient accretion flow (RIAF) model,
which are proposed as models for the accretion flow with
higher ($\sim L_{\rm E}$) and lower luminosities, respectively
(\citeauthor{SS73} \citeyear{SS73}; 
\citeauthor{Ichi77} \citeyear{Ichi77}; 
\citeauthor{Rees82} \citeyear{Rees82}; 
\citeauthor{Abramo88} \citeyear{Abramo88}; 
\citeauthor{NY94} \citeyear{NY94}, 
see \citeauthor{KFM08} \citeyear{KFM08} for an extensive review).
These models have good prediction powers and thus make it possible to
directly compare the theory with the observations 
(e.g., through spectral fitting),
however, the results derived those simplified models need to be checked,
since their results may depend on the $\alpha$-viscosity assumption
and radially one-dimensional approximation.
In fact, \citet{Hirose09} have claimed that the 
radiation-pressure-dominated 
part of the disk is thermally stable, though it was shown to be
unstable under the $\alpha$-viscosity prescription
\citep{Shiba75, SS76}.

As a distinct line of disk research,
multi-dimensional hydrodynamical flow simulations were attempted
based on the $\alpha$-viscosity prescription 
\citep{Igumen99,Stone99,MG02},
but it was not a main stream of research. 
After the identification of the disk viscosity,
the situations drastically changed.
Global magnetohydrodynamic (MHD) simulations have been 
extensively performed by several groups
(\citeauthor{Matsumoto99} \citeyear{Matsumoto99}; 
\citeauthor{MHM00} \citeyear{MHM00}; 
\citeauthor{HK01} \citeyear{HK01}; 
\citeauthor{Koide01} \citeyear{Koide01}; 
\citeauthor{DV03} \citeyear{DV03}, 
\citeauthor{HK06} \citeyear{HK06}, 
see, however, a review by 
\citeauthor{Spruit10} \citeyear{Spruit10}).
Despite these studies, 
there is no wide consensus regarding the global and local behavior 
of magnetic fields; e.g.,
it is an open question
how numerical results are sensitive to the initial magnetic field
strengths and configurations,
boundary conditions, and
numerical resolutions.
This is due partly to highly nonlinear spatio-temporal 
evolution of magnetic fields in differentially rotating media.
Further, most of the previous global MHD simulations are non-radiative ones
and, hence, they cannot model accretion disks in high-luminosity states,
in which significant radiative cooling (and occasionally 
strong matter-radiation coupling) is expected.  
The radiation transfer should be solved to explain
the energy release processes within the disk. 
The dynamical effects of radiation are especially important 
for the radiation-pressure-dominated disk, 
since they are expected to produce strong radiatively driven outflows,
thereby the disk structure being significantly modified. 


In \citet{O09} we have presented a new type of accretion flow
simulation based on the global radiation-magnetohydrodynamic (RMHD)
simulations. We solved the problem from the first principle
(i.e., without employing the phenomenological $\alpha$-viscosity prescription),
including the case of radiatively very efficient accretion flow.
That is, we considered the following basic processes in the accretion flow and jets:
the transport of angular momentum induced via the magnetic torque, 
leading to the accreting motion,
the conversion of the mechanical energy to the thermal energy via the MHD processes,
the dissipation of thermal energy, the radiative transfer,
and radiation-pressure and Lorentz forces,
which play important roles in launching outflows and supporting the disks
in the vertical direction.
The overview of the RMHD simulation results of the accretion flow 
and outflow around the black holes has already been published 
elsewhere \citep[e.g.,][]{O09}. 
Here, we present the detailed analyses of the simulated flow 
structure based on our newly performed, improved simulations.

Similar RMHD simulations were attempted previously
but only under the shearing-box approximation, 
in which only a local patch of the disk is treated
\citep[e.g.,][]{Turner03,Hirose06}. 
Global (radial) coupling of magnetic fields are ignored in those
simulations and thus collimated outflows cannot be produced there.
Further, advective motion of gas and photons were not considered.
For establishing a realistic picture of accretion disks and outflow,
therefore, global, multi-dimensional RMHD simulations, 
of a kind reported in the present paper, are indispensable.
We will show that the realistic flow properties simulated here
share some similarities with the flows described by
the previous one-dimensional models but that they also
exhibit new features which were not anticipated previously.
The plan of this paper is as follows.
Basic equations and physical assumptions are explained in Section 2,
and numerical procedures are described in Section 3.
We will then show the results of simulations in Section 4 
and give discussion in Section 5. 
Finally, Section 6 is devoted to summary.

\section{Basic Equations and Assumptions}
\label{basic}
We solve a full set of RMHD equations
under flux-limited diffusion (FLD) approximation
in the cylindrical coordinates, $(r,\varphi,z)$.
In the present study, we assume that the flow is non-self-gravitating,
reflection symmetric relative to the equatorial plane (with $z=0$), 
and axisymmetric with respect to the rotation axis 
(i.e., $\partial/\partial\varphi=0$). 
We describe the gravitational field of the black hole in terms of 
pseudo-Newtonian hydrodynamics, in which the gravitational potential 
is given by $\psi=-GM/(R-R_{\rm S})$ \citep{PW80},
where $R [\equiv (r^2+z^2)^{1/2}]$ is the distance from the origin 
and $R_{\rm S}(\equiv 2GM/c^2)$ is the Schwarzschild radius
(with $M$ and $c$ being the black hole mass and the light velocity,
respectively).

The basic equations are the continuity equation,
\begin{equation}
  \frac{\partial \rho}{\partial t}
  + {\bma \nabla} \cdot (\rho {\bma v}) = 0,
  \label{mass_con}
\end{equation}
the equations of motion,
\begin{eqnarray}
  \frac{\partial (\rho {\bma v})}{\partial t}
  + {\bma \nabla} \cdot 
  \left( \rho {\bma v}{\bma v} - \frac{{\bma B}{\bma B}}{4 \pi} \right) 
  = - {\bma \nabla} \left( p+\frac{|{\bma B}|^2}{8\pi} \right) \nonumber \\
  +\frac {\chi}{c} {\bma F}_0
  -\rho{\bma \nabla}\psi,
  \label{mom}
\end{eqnarray}
the energy equation of the gas,
\begin{equation}
  \frac{\partial e }{\partial t}
   + {\bma \nabla}\cdot(e {\bma v}) 
   = -p{\bma \nabla}\cdot{\bma v} 
   +\frac{4\pi}{c^2}\eta J^2
   -4\pi \kappa B + c\kappa  E_0,
   \label{gase}
\end{equation}
the energy equation of the radiation,
\begin{equation}
  \frac{\partial E_0}{\partial t}
   + {\bma \nabla}\cdot(E_0 {\bma v}) 
   = -{\bma \nabla}\cdot{\bma F_0} -{\bma \nabla}{\bma v}:{\bma {\rm P}_0}
  + 4\pi \kappa B - c\kappa E_0,
  \label{rade}
\end{equation}
and the induction equation,
\begin{equation}
 \frac{\partial \bma B}{\partial t}
  ={\bma \nabla} \times 
  \left({\bma v}\times {\bma B}-\frac{4\pi}{c}\eta{\bma J} \right).
  \label{ind}
\end{equation}
Here, $\rho$ is the gas mass density,
$\bma{v}=(v_r, v_\varphi, v_z)$ is the velocity,
$e$ is the internal energy density of the gas,
$p$ is the gas pressure,
$\bma{B}=(B_r, B_\varphi, B_z)$ is the magnetic field,
$\bma{J}(=c{\bma \nabla}\times {\bma B}/4\pi)$ 
is the electric current,
$\eta$ is the resistivity,
$B$ is the blackbody intensity,
$E_0$ is the radiation energy density,
${\bma F}_0$ is the radiation flux,
${\bma {\rm P}}_0$ is the radiation pressure tensor,
$\kappa$ is the absorption opacity,
and $\chi$ is the total opacity.

The gas pressure is related to the internal energy density of the gas 
by
\begin{equation}
  p=(\gamma-1)e,
\end{equation}
where $\gamma$ is the specific heat ratio.
The temperature of the gas, $T_{\rm gas}$, can then be calculated from
\begin{equation}
  p=\frac{\rho k T_{\rm gas}}{\mu m_{\rm p}},
\end{equation}
where $k$ is the Boltzmann constant, 
$\mu$ is the mean molecular weight, 
and $m_{\rm p}$ is the proton mass.

For the absorption opacity,
we consider the Rosseland mean free$-$free absorption, $\kappa_{\rm ff}$, 
and bound$-$free absorption for solar metallicity, $\kappa_{\rm bf}$,
\begin{equation}
  \kappa=\kappa_{\rm ff}+\kappa_{\rm bf},
\end{equation}
where $\kappa_{\rm ff}$ and $\kappa_{\rm bf}$
are given by
\begin{equation}
  \kappa_{\rm ff} = 1.7\times 10^{-25} T^{-7/2}
  \left(\frac{\rho}{m_{\rm p}}\right)^2 \rm cm^{-1},
\end{equation}
\citep{RL79}, and 
\begin{equation}
  \kappa_{\rm bf} = 4.8\times 10^{-24} T^{-7/2}
  \left(\frac{\rho}{m_{\rm p}}\right)^2
  \rm cm^{-1},
\end{equation}
\citep{HHS62}.
The total opacity is given by
\begin{equation}
 \chi=\kappa+\frac{\rho \sigma_{\rm T}}{m_{\rm p}},
\end{equation}
with $\sigma_{\rm T}$ being the Thomson scattering cross-section.

We employ the FLD approximation
developed by \citet{LP81} 
(see also \citeauthor{TS01} \citeyear{TS01}).
This assumption is valid in both the optically thick diffusion limit 
and the optically thin free-streaming limit in one-dimensional space.
In this framework,
the radiation flux is expressed
in terms of the gradient of the radiation energy density via
\begin{equation}
  {\bma F}_0 = -\frac{c\lambda}{\chi}\nabla E_0.
   \label{FLDflux}
\end{equation}
Here, the dimensionless function ($\lambda$),
which is called the flux limiter, is given by
\begin{equation}
  \lambda = \frac{2+{\cal R}}{6+3{\cal R}+{\cal R}^2},
\end{equation}
using the dimensionless quantity, 
${\cal R}=\left| \nabla E_0 \right| / \left( \chi E_0 \right)$.
The radiation pressure tensor is written as 
\begin{equation}
 {\bma {\rm P}}_0 = {\bma {\rm f}} E_0.
  \label{tensor}
\end{equation}
Here, ${\bma {\rm f}}$ is the Eddington tensor
and its components are
\begin{equation}
  {\bma {\rm f}} = \frac{1}{2}(1-f){\bma {\rm I}}
  +\frac{1}{2} (3f-1) {\bma n}{\bma n},
\end{equation}
where $f$ is the Eddington factor,
\begin{equation}
  f = \lambda + \lambda^2 {\cal R}^2,
\end{equation}
and ${\bma n}$ is the unit vector in the direction of the radiation
energy density gradient,
\begin{equation}
  {\bma n} = \frac{\nabla E_0}{\left| \nabla E_0 \right|}.
\end{equation}
We find
$\lambda \rightarrow 1/3$ and $f \rightarrow 1/3$
due to ${\cal R} \rightarrow 0$
in the optically thick limit. 
On the other hand, 
in the optically thin limit of ${\cal R}\rightarrow \infty$,
we have $| F_0 | = cE_0$.
These give correct relations in the optically thick diffusion limit
and the optically thin streaming limit, respectively.

Throughout the present study,
we assume $M=10M_{\odot}$, $\gamma=5/3$, and $\mu=0.5$.
The resistivity is assumed to be constant, $\eta=10^{-3}c R_{\rm S}$,
although we have employed the anomalous resistivity \citep{Yokoyama94}
in the previous paper \citep{O09}.
Note that our results do not change so much 
by switching the resistivity.
The uniform resistivity might have an advantage over 
the anomalous resistivity in longer simulations, 
since it works to dissipate the magnetic fields and 
prevent the Alfv\'en speed from being too large. 
Two-dimensional non-radiative MHD simulations 
using a uniform resistivity
have been performed by \citet{KHM04}.
They succeeded in reproducing the semirelativistic jets from the 
accretion disks.

Here we stress that in our RMHD simulations
the magnetic torque 
is responsible for the angular-momentum transfer 
and the Joule heating for the heating of the matter.
In the conventional disk models 
and in the radiation hydrodynamic simulations,
in contrast, the phenomenological viscosity 
($\alpha$-viscosity) induces the angular-momentum transfer 
as well as the energy dissipation
\citep{ECK88, Okuda00, O05b, O06}.
The radiative cooling and radiation-pressure force are both considered 
in the present study, while they cannot be taken into account
in the non-radiative MHD simulations.

\section{Numerical Methods and Models}
\subsection{Outline}
Before presenting detailed calculation methods
we outline the procedure of our simulations in this subsection.
We solve the time evolutions of a torus by solving the
basic radiation-MHD equations under the assumption
of the FLD (see Section 2).
The initial torus is in hydrostatic balance
and is surrounded by a hot rarefied atmosphere (see Section 3.3).
We let the initial torus evolve by solving non-radiative MHD equations,
for the first elapsed time of $t \sim 1$ s,
which corresponds to $\sim 4.5$ times the Keplerian time at the center of
the torus (at $r_0 = 40 R_{\rm S}$). We then turn on the radiative terms
in the basic equations and further solve the evolution of the torus for
$t\sim 10 $ s.
Whereas the density normalization (which is the central density of
the initial torus in the present study)
can be taken arbitrarily in the non-radiative MHD
simulations,
simulations with the radiative terms do depend on
the density normalization. That is, the density normalization controls
the relative importance of the radiative cooling.
We will be able to reproduce three distinct modes of accretion flow
by changing the density normalizations (see Section 3.4).


\subsection{Code}
\label{code}
We numerically solve the set of RMHD equations 
using an explicit-implicit finite difference scheme.
The time step is restricted by the Courant$-$Friedrichs$-$Levi condition.
The numerical procedure is divided into the following steps.
%
In step I,
the MHD terms are solved by the modified Lax$-$Wendroff scheme 
\citep{RB67}.
Equations (\ref{mass_con})$-$(\ref{gase}) and (\ref{ind}),
except the gas$-$radiation interaction terms of Equation (\ref{gase}), 
are solved in this step.
The numerical procedure in this step is basically the same as that used 
in \citet{KHM04},
except that the equation of the internal energy of gas is
solved in the present simulations,
while they calculated the evolution of the total energy 
(internal and kinetic energies of the gas and magnetic energy).
In step II, an integral formulation is used to generate
a conservative differencing scheme for the advection term of
Equation (\ref{rade}).
Steps I and II are solved with the explicit method, 
while steps III and IV are solved with the implicit method.
The radiation energy and gas energy are updated simultaneously
via the gas-radiation interaction in step III.
We consider
the third and final terms of the right-hand side of Equation (\ref{gase})
as well as the terms on the right-hand side of Equation (\ref{rade})
except for the radiative flux term in this step.
In the final step (step IV),
we update the radiation energy density via 
the radial radiative flux
(the first term of the right-hand side of Equation (\ref{rade})).
The radiation energy density is advanced again 
by the vertical radiative flux.
In this step, the Thomas method is used for matrix conversion. 

Our method of calculation is an extension of those of MHD 
simulations \citep[e.g.,][]{KHM04}. 
The MHD code was tested by the problems of the propagation 
of a two-dimensional MHD wave and of a one-dimensional shock, 
and applied to the simulations of the magnetized disks and jets
\citep[see also][]{KMS04}.
We also performed the tests of gas$-$radiation interaction, 
one-dimensional and two-dimensional radiation front propagation, 
and sub- and super-critical shock, and radiation-dominated shock, 
finding that the results are in good agreement with those 
\citet{TS01}.
%

\subsection{Initial Conditions}
\label{IC}
We initially set a rotating torus in hydrostatic balance.
We employ a polytropic equation of state,
$p=\rho^{1/n}$ with $n=3$
and a power-law-specific angular-momentum distribution,
$l=l_0(r/r_0)^a$ with 
$l_0=(GMr_0^3)^{1/2}/(r-R_{\rm S})$, $r_0=40R_{\rm S}$, and $a=0.46$.
The initial density and gas pressure distributions are expressed as 
\begin{equation}
 \rho_t(r, z)=\rho_0 
  \left[ 
   1-\frac{1}{\gamma \epsilon_0^2\psi^2(r_0,0)}
   \frac{\psi_{\rm eff}(r, z)-\psi(r_0, 0)}{n+1}
  \right]^n,
\end{equation}
and 
\begin{equation}
 p_t=\rho_0\gamma\epsilon_0^2\psi^2(r_0,0)
  \left[
   \frac{\rho_t(r, z)}{\rho_0}
  \right]^{1+1/n},
\end{equation}
where $\rho_0$ is the initial density at the center of torus,
the parameter, $\epsilon_0$, is set to be $1.45\times 10^{-3}$,
and $\psi_{\rm eff}$ is the effective potential given by
$\psi_{\rm eff}(r, z) = \psi(r, z)+0.5(l/r)^2/(1-a)$.

The initial magnetic fields in the torus are purely poloidal
($B_\varphi=0$). 
Their distribution is described in terms of the azimuthal component 
of the vector potential, which is assumed to be proportional 
to the density, $A_\varphi\propto\rho_t$.
Other components are set to be zero, $A_r=A_z=0$.
We initially set the plasma-$\beta$, 
the ratio of gas pressure to magnetic pressure, 
to be $100$. 

This initial torus is embedded in a nonrotating, hot, 
and rarefied corona with no magnetic fields.
The density and pressure distributions of the corona are
\begin{equation}
 \rho_c(r, z)=\rho_1
  \left[
   -\frac{\psi(r, z)+GM/R_{\rm S}}{\epsilon_c GM/R_{\rm S}}
  \right]
\end{equation}
and 
\begin{equation}
 p_c(r, z)=\rho_c(r, z)\frac{\epsilon_c GM}{R_{\rm S}},
\end{equation}
with $\rho_1=10^{-6}\rho_0$ and $\epsilon_c=1.0$.
This corona is initially in hydrostatic equilibrium.
Our initial conditions are the same as those of model B in \citet{KMS04},
except the density of the corona is smaller by a factor of 20.

\subsection{Models and Grids}
We need to assign the density normalization
when starting RMHD simulations.
In total,
we calculate three models in the present study with
$\rho_0=1\, \rm g \, cm^{-3}$ (model A),
$10^{-4}\, \rm g \, cm^{-3}$ (model B),
and $10^{-8}\, \rm g \, cm^{-3}$ (model C).
We will show in Section 4 that our three models with high, moderate, and low
density normalizations correspond to the slim disk,
the standard disk, and the RIAF models, respectively.
Note that we set the lower limit of the density to be 
$\rho/\rho_0=10^{-10}$ for all models. 
Such a lower limit sometimes appears in the upper regions
($z>60Rs$) around the rotation axis in model A, 
since the outflowing velocity is so large. 
In models B and C, the density is rarely below the limit.

Since we assume axisymmetry around the rotation axis, 
as well as reflection symmetry
with respect to the equatorial plane, 
the computational domain can be restricted to one quadrant 
of the meridional plane.
For models A and C, 
the number of grid points is $(N_r, N_z)=(512, 512)$.
The grid spacing which is uniform ($\Delta r=\Delta z=0.2R_{\rm S}$),
extends from $2R_{\rm S}$ to $105R_{\rm S}$ in the radial direction 
and from $0$ to $103R_{\rm S}$ in the vertical direction. 
In model B,
on the other hand, 
we use smaller spacing grids, $\Delta r=\Delta z=0.1R_{\rm S}$,
since the scale height of the disk is smaller (see below).
The number of grid points is $(N_r, N_z)=(1024, 512)$,
and the computational domain is $2R_{\rm S}\leq r \leq 105R_{\rm S}$ and 
$0 \leq z\leq 51R_{\rm S}$.

\subsection{Boundary Conditions}
For the matter and magnetic fields,
we adopt free boundary conditions
at the inner and outer boundaries ($r=2R_{\rm S}$ and $103R_{\rm S}$) 
and upper boundary 
($z=103R_{\rm S}$ for models A and C, $z=51R_{\rm S}$ for model B).
That is, the matter can freely go out but not enter 
and the magnetic fields do not change across 
the inner, outer, and upper boundaries.
If the radial component of the velocity is negative (or positive) 
at the outer (or inner) boundary, 
it is automatically set to be zero. 
The vertical component of the velocity is also set to be zero, 
when it is negative at the upper boundary. 
With respect to the disk plane ($z=0$), 
we assume that $\rho$, $p$, $v_r$, $v_\varphi$, and $B_z$, 
and are symmetric, 
while $v_z$, $B_r$, and $B_\varphi$ are antisymmetric.

The radiative fluxes are assumed 
without using a gradient of the radiation energy density
at the inner, outer, and upper boundaries.
The vertical and radial components of the radiative fluxes
are set to be $c E_0$ at the outer and upper boundaries. 
At the inner boundary ($r=2R_{\rm S}$), 
we set the radial component of the radiative flux 
to be zero except at $0<z<2R_{\rm S}$. 
Since we also set the radial component of the advection 
of the radiation energy to be zero,
the radiation energy neither leaves nor enters.
At the boundary at $r=2R_{\rm S}$ with $0<z<2R_{\rm S}$, 
we assume $F_0^r=-cE_0$, 
meaning that the radiation is swallowed by the black hole.
In addition, we have $F_0^z=0$ at $z=0$, 
since we impose a symmetric boundary condition at the equatorial plane.


\subsection{Prelusive Calculation}
\begin{figure}[h]
 \epsscale{0.85}
 \plotone{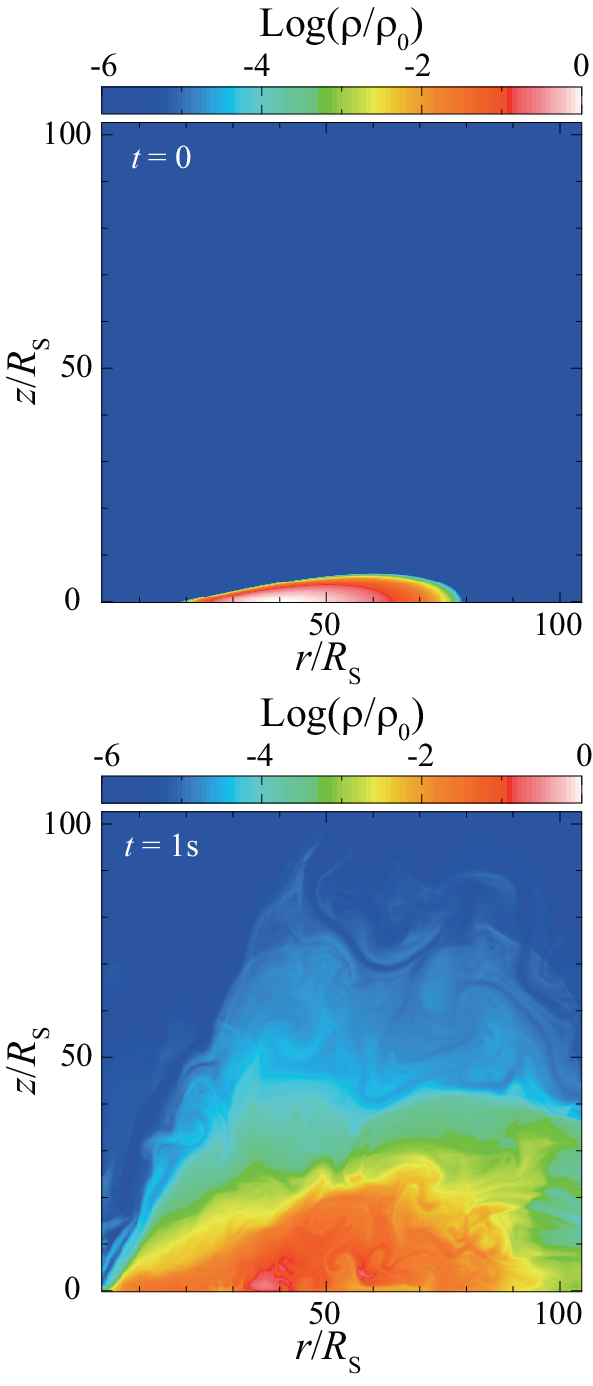}
 \caption{Color contour of the initial matter density distribution (top)
 and that obtained after 1 s non-radiative MHD simulations (bottom).
 \label{figIC}   
 }
\end{figure}
As we have already mentioned in Section 3.1,
we evolve the initial torus by solving non-radiative MHD equations
for 1 s.
Then, we solve only the MHD terms with gravity using step I
in Section \ref{code}.
In Figure \ref{figIC}, we show 
the density distribution of the initial torus (top panel).
Although the initial torus is in hydrostatic balance (see Section \ref{IC}),
the matter falls toward the black hole 
since the angular momentum is transported by the magnetic torque.
Moreover, the magnetic energy dissipates and the flow is heated up.
Thus, the initial torus also expands in the vertical direction.
The resulting density distribution at $t=1$ s
is shown in the bottom panel of Figure 1.
Here, $512 \times 512$ grids (the grid spacing is $0.2R_{\rm S}$) are used.
We have confirmed that 
the resulting density distribution does not change so much,
even if we employ smaller mesh spacings,
$\Delta r=\Delta z=0.1R_{\rm S}$ ($1024 \times 512$ grids).
Assigning the density normalization,
and setting the radiation temperature to be $10^4$ K
in the whole region,
we start the RMHD simulations 
from the resulting structure given by the prelusive MHD calculations 
and go on performing them until $t\sim 10$ s.

\subsection{Mass Accretion Rates, Outflow Rates, and Luminosities}
\label{rates}

The mass accretion rate is calculated 
at the inner boundary at $r=2R_{\rm S}$ by
\begin{equation}
 \dot{M}_{\rm acc}=-2 \int_{0}^{2R_{\rm S}} 2\pi r \rho v_r dz,
\end{equation}
which is the mass passing through the inner boundary 
per unit time.
The photon luminosity is calculated by 
\begin{equation}
 L_{\rm ph}=2\int_{2R_{\rm S}}^{r_{\rm c1}} 2\pi r 
  \left(F^z_0+v_z E_0 \right) dr,
  \label{Lph}
\end{equation} 
at the height of $z=z_{\rm c}$.
The values of $(r_{\rm c1}, z_{\rm c})$ 
are carefully chosen so as not to include the contribution from
the initial torus.
We, hence, set 
$(r_{\rm c1}, z_{\rm c})=(25R_{\rm S}, 60R_{\rm S})$ for model A,
$(30R_{\rm S}, 30R_{\rm S})$ for model B,
and $(60R_{\rm S}, 60R_{\rm S})$ for model C.
In model C we employ relatively larger $r_{\rm c1}(=60R_{\rm S})$,
however, the contribution to the total emission from the outer region 
is very small, since the density is very small there.
The mass outflow rate and the kinetic luminosity
are also calculated at $z=z_{\rm c}$ as
\begin{equation}
 \dot{M}_{\rm out}=2\int_{2R_{\rm S}}^{r_{\rm c2}} 2\pi r 
  \rho v_z
  H\left(v_R-v_{\rm esc} \right) dr,
\end{equation}
\begin{equation}
 L_{\rm kin}=2\int_{2R_{\rm S}}^{r_{\rm c2}} 2\pi r 
  \left(\frac{1}{2}\rho v_R^2 \right)
  v_z H\left(v_R-v_{\rm esc} \right) dr,
\end{equation}
where $H$ is the Heaviside step function; 
i.e., $H(x)=1$ for $x\ge 0$ 
and $H(x)=0$ for $x<0$,
and $v_R[\equiv v_r(r/R)+v_z(z/R)]$ 
is the $R$-component of the velocity.
We employ $(r_{\rm c2}, z_{\rm c})=(60R_{\rm S}, 60R_{\rm S})$
for models A and C, and
$(r_{\rm c2}, z_{\rm c})=(30R_{\rm S}, 30R_{\rm S})$ for model B.
That is, what are meant by $\dot{M}_{\rm out}$ and $L_{\rm kin}$
in the present study indicate the mass and kinetic energy transported 
upward only
by the high-velocity outflow ($v_R>v_{\rm esc}$) per unit time.
In addition, we plot the luminosity of trapped radiation,
which is evaluated by
\begin{equation}
  L_{\rm trap}=-2 \int_{0}^{2R_{\rm S}} 2\pi r 
   \left(F_0^r+v_r E_0 \right) dz,
\end{equation}
with $r=2R_{\rm S}$.
It implies the radiation energy
swallowed by the black hole per unit time.

\section{Results}
\label{results}
We first overview the simulation results in Section \ref{overview}
and then provide more detailed information
in Sections \ref{modelA}$-$\ref{modelC}
for models A, B, and C, respectively.

\subsection{Overview of Simulated Flows}
\label{overview}
\begin{figure*}
 \epsscale{1.18}
 \plotone{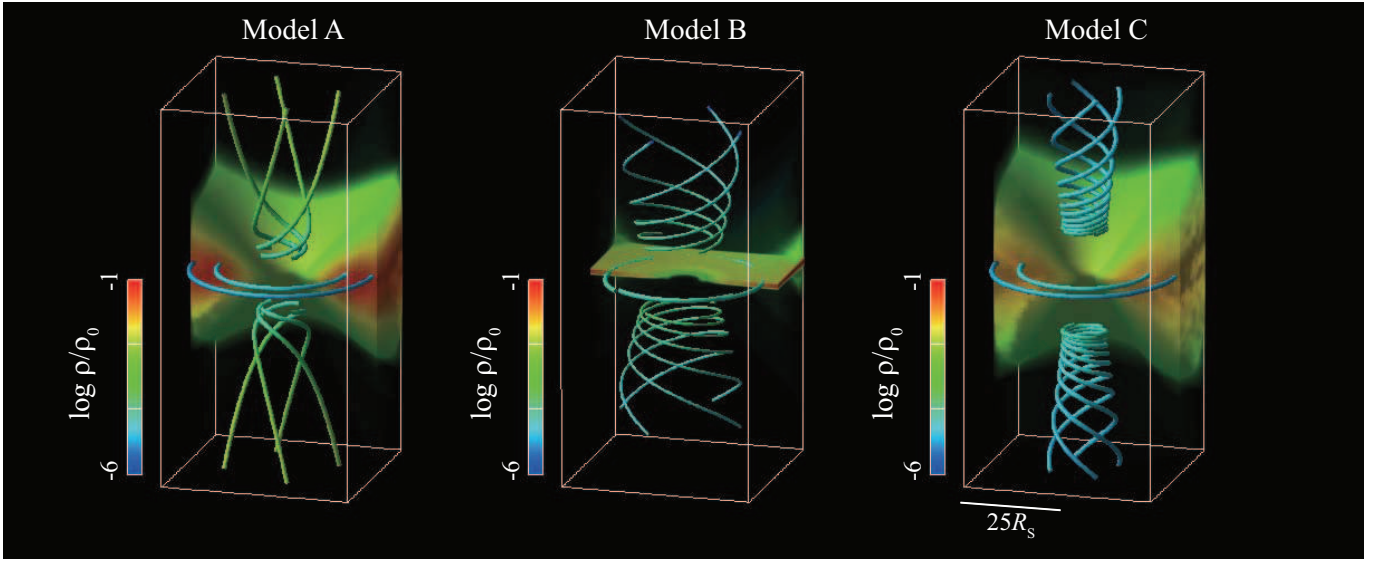}
 \caption{
 Perspective view of inflow and outflow patterns near the black hole
 for models A, B, and C, from left to right, respectively. 
Also plotted are
 normalized density distributions (color) and streamlines,
 which are time-averaged over 6$-$7 s for models A and C
 and over 9$-$10 s for model B.
 \label{figview}   
 }
\end{figure*}
\begin{figure*}
 \epsscale{0.68}
 \plotone{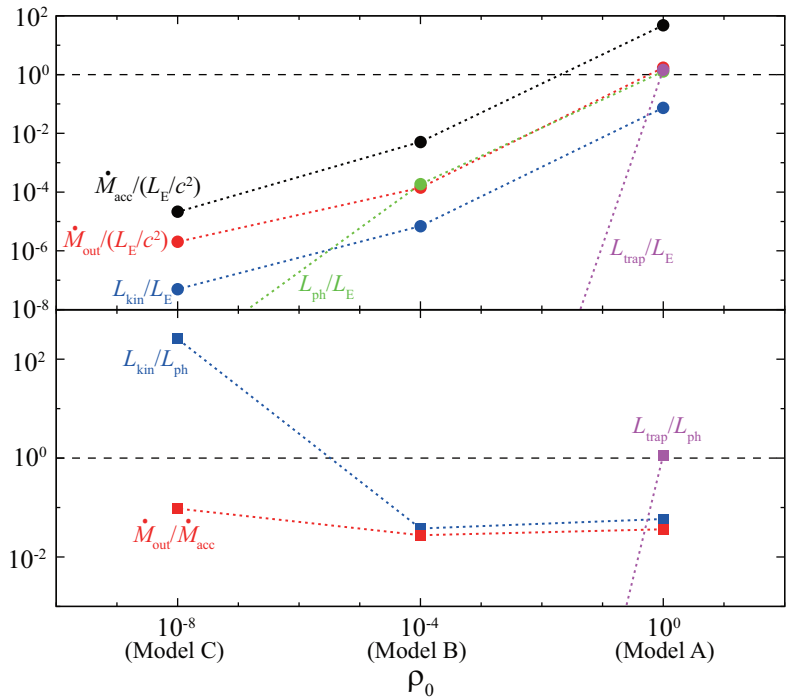}
 \caption{
 Top panel:
 mass accretion rate, $\dot{M}_{\rm acc}$, 
 and mass outflow rate, $\dot{M}_{\rm out}$, 
 normalized by the critical rate, $L_{\rm E}/c^2$, 
 and the photon, kinetic, and trapping luminosities
 normalized by the Eddington luminosity, $L_{\rm E}$.
 Bottom panel:
 the ratios of $\dot{M}_{\rm out}/\dot{M}_{\rm acc}$,
 $L_{\rm kin}/L_{\rm ph}$, and $L_{\rm trap}/L_{\rm ph}$.
 All values are time-averaged over 5$-$7.5 s (models A and C)
 and 7.5$-$10 s (model B).
 \label{figML}   
 }
\end{figure*}
\begin{figure*}
 \epsscale{0.75}
 \plotone{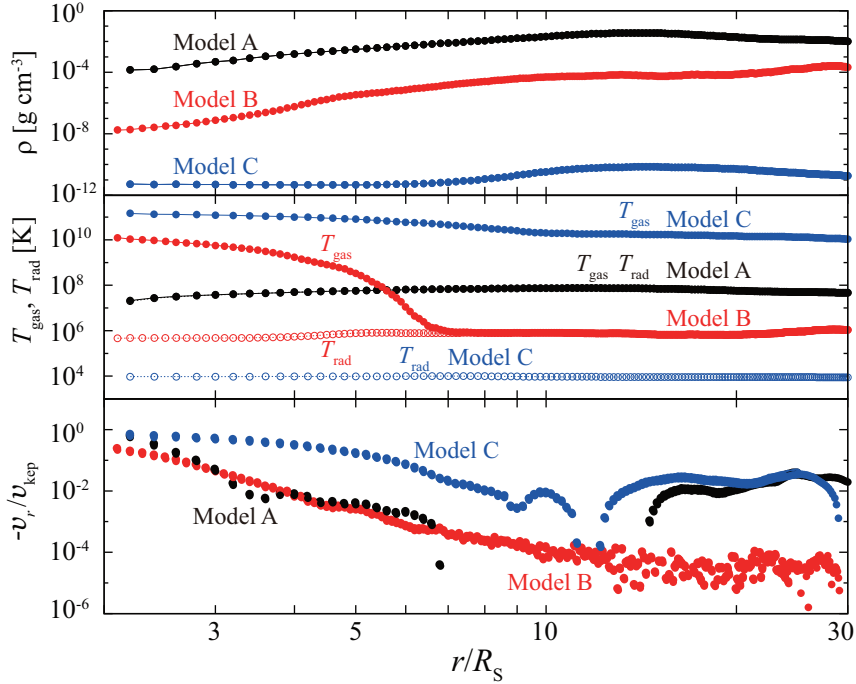}
 \caption{
 Radial profiles of the density (top), 
 the gas and radiation temperatures (middle),
 and the inflow velocity normalized
 by the local Keplerian velocity (bottom) 
 near the equatorial plane
 for models A ($z=0.1R_{\rm S}$, black), 
 B ($z=0.05R_{\rm S}$, red), 
and C ($z=0.1R_{\rm S}$, blue).
 All values are time-averaged over 7.5$-$10 s.
 \label{figmid}   
 }
\end{figure*}

The overall flow structures obtained by our RMHD simulations
can be summarized in Figures \ref{figview}--\ref{figmid}.
We first show perspective views of simulated flows in models A$-$C
in Figure \ref{figview}.  Here, the color contours 
indicate the distributions of normalized density, $\rho/\rho_0$, 
time-averaged over 6$-$7 s for models A and C
and over 9$-$10 s for model B.
We find that a geometrically thick disk forms in models A and C,
while a geometrically thin disk forms in model B.
The streamlines indicated by the thick lines are overlaid in this figure.
We find in all models that the gas near the equatorial plane 
is on a quasi-circular orbit around the central black hole, 
whereas the gas away from the equatorial plane 
shows helical and outflowing motion.
The helical motion means that the outflow material
processes a substantial amount of angular momenta.

The different dynamical properties of accretion flows in three models 
are more quantitatively shown in Figure \ref{figML}.
In the top panel of the figure,
we plot the normalized mass accretion rate, 
$\dot{M}_{\rm acc}/(L_{\rm E}/c^2)$,
mass outflow rate, $\dot{M}_{\rm out}/(L_{\rm E}/c^2)$,
photon luminosity, $L_{\rm ph}/L_{\rm E}$,
kinetic luminosity, $L_{\rm kin}/L_{\rm E}$, 
and trapping luminosity, $L_{\rm trap}/L_{\rm E}$. 
We also plot $L_{\rm kin}/L_{\rm ph}$, $L_{\rm trap}/L_{\rm ph}$,
and $\dot{M}_{\rm out}/\dot{M}_{\rm acc}$ in the bottom panel.
They are time-averaged over $t=5$$-$7.5 s for models A and C
and over 7.5$-$10 s for model B.
For the calculation methods of these quantities, see Section \ref{rates}.

We find in model A that 
the photon luminosity exceeds the Eddington luminosity
and that the trapping luminosity is substantial,
implying that the model A flow is supercritical flow.
The thin disk in model B corresponds to the standard-type disk,
since low scaleheight is a result of efficient radiative cooling.
By contrast, the photon luminosity is negligible in model C,
indicating that the model C flow is RIAF.
We also see 
a large value of $L_{\rm kin}/L_{\rm ph}$ $(>1)$ only in model C.

In Figure \ref{figmid},
we show the radial profiles of the density (top), 
the gas and radiation temperatures (middle),
and the radial velocities (bottom) around the equatorial plane.
They are time-averaged over $t=7.5$$-$10 s.
As expected, the gas temperature is highest in model C
(RIAF), while it is lowest in model B except at radii
less than $\sim 5R_{\rm S}$.
The decoupling of the gas and radiation temperatures occur
in model C entirely and in model B, inside $\sim 7R_{\rm S}$.

To summarize, we could reproduce three distinct states
of accretion flow with the same code but 
by changing the density normalization.
In the subsequent subsections
we give more information individually for models A$-$C.

\subsection{Model A}
\label{modelA}
\subsubsection{Overview}
\label{overA}
Model A, with a high density normalization 
($\rho_0=1.0\, \rm g\, cm^{-3}$),
corresponds to the two-dimensional
RMHD version of the slim disk model
\citep{Abramo88, Wat+00}
but with significant outflow \citep{TOM09}.
The geometrically thick disk
is supported by the radiation pressure
(see Section \ref{2DA} and Section \ref{VerA}).
We also find in Figure \ref{figmid}
that the disk consists of very dense 
and moderately hot ($10^{7-8}$ K) matter.
The gas temperature is comparable to the radiation temperature
due to the effective gas$-$radiation interaction.
The inflow velocity ($-v_r$) tends to increase inward only
in the vicinity of the black hole, $r\lsim 7R_{\rm S}$,
and it is very small or even negative (i.e., outflow) 
around $r\sim 10R_{\rm S}$
(see the bottom panel in Figure \ref{figmid}).
Basic properties of model A are
roughly consistent with those of the slim disk.
However, the slopes of the radial profiles 
of the density and temperatures do
largely differ from those of the slim disk.
(Note that such deviations from the conventional disk models
are also found in models B and C.)
These discrepancies might be due to the limited calculation time,
the restricted computational domain,
and the influence of the initial conditions.
Detailed comparison is left for future work.

In Figure \ref{figML},
we find that the mass accretion rate onto the black hole
is much larger than the critical value,
$\dot{M}_{\rm acc} \sim 10^2 L_{\rm E}/c^2$.
The photon luminosity is $L_{\rm ph}\sim 1.7 L_{\rm E}$,
which is calculated 
based on the vertical component of the radiative flux within 
the polar angle of $\theta=\tan^{-1}(25R_{\rm S}/60R_{\rm S})\sim 23^\circ$.
(Note that we assigned $(r_{\rm c1}, z_{\rm c})=(25R_{\rm S}, 60R_{\rm S})$
in Equation (\ref{Lph}).)
Since $(r_{\rm c1}, z_{\rm c})=(25R_{\rm S}, 60R_{\rm S})$ lies on 
the line of the photosphere, $z/r\sim 2.4$ (see Section \ref{2DA}),
the photon luminosity calculated with these values is 
equal to the sum of 
the radiation energy released at the photosphere of 
the inner part of the disk, $r<25R_{\rm S}$, per unit time.
Also, since $r_{\rm c1}=25R_{\rm S}$ is smaller than 
the position of the initial torus, $r=40R_{\rm S}$,
we can avoid possible influence by the initial conditions.
Here we note that the calculated luminosity ($L_{\rm ph}$)
may be underestimated, since 
the contribution from the outer disk ($r\gsim 25R_{\rm S}$)
was ignored.
For better calculations of $L_{\rm ph}$,
we need simulations with a larger computational domain,
in which we set the initial torus at a much larger radius.

In addition, 
the photons mainly escape along the rotation axis,
producing the mildly collimated emission
(note that the vertical component of the radiative flux
is much larger than the radial one around the rotation axis, $z/r>2.4$).
Thus, the apparent luminosity of the flow would exhibit
the strong viewing-angle dependence.
We will discuss this issue in Section \ref{discussions}.

Figure \ref{figML}
also indicates that the trapping luminosity 
is slightly larger than the Eddington luminosity (top panel),
and is comparable to the photon luminosity (bottom panel).
This means that the large number of photons generated inside the
disk is swallowed by the black hole together with accreting matter,
without many being released from the disk surface.

Although we do not represent 
the rotation velocity in Figure \ref{figmid},
the matter approximately rotates 
with the Keplerian velocity 
(within 10\%) near the equatorial plane.
The outflow is driven by the radiation force
(see Sections \ref{2DA} and \ref{VerA}).
The mass-outflow rate is a few \% of the mass accretion rate
(see Figure \ref{figML}).
We also find $L_{\rm kin}\lsim 0.1L_{\rm ph}$ in this figure.
This implies that the supercritical flows in model A
release energy via radiation rather than via outflows.

\subsubsection{Two-dimensional Structure}
\label{2DA}
\begin{figure*}
 \epsscale{0.825}
 \plotone{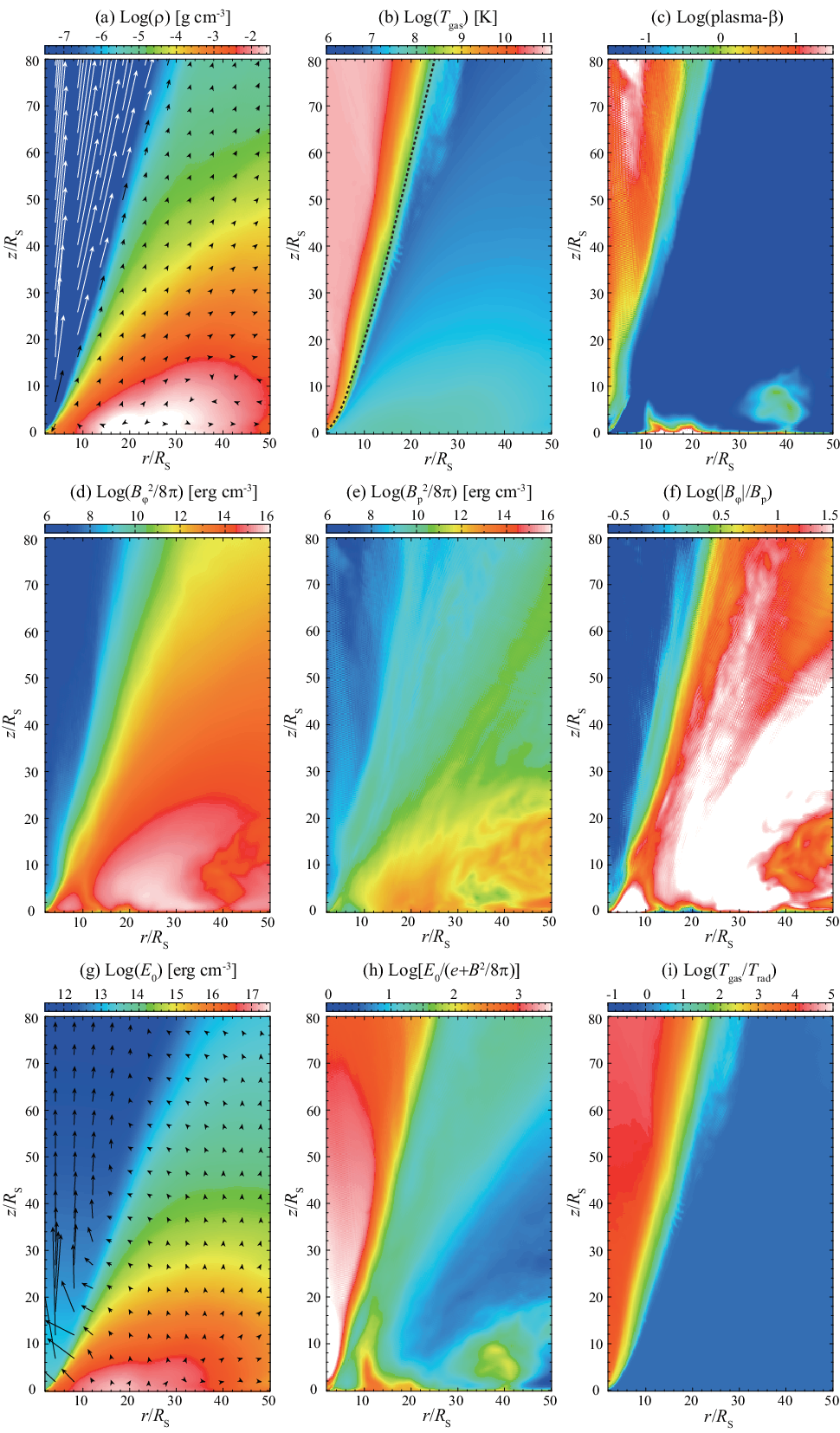}
 \caption{Two-dimensional distribution of
 the various quantities for Model A:
 (a) the density overlaid with the velocity vectors, 
 (b) the gas temperature, 
 (c) the plasma-$\beta$,
 (d) the magnetic energies via the toroidal component of field,
 (e) the same but of the poloidal component,
 (f) the magnetic pitch,
 (g) the radiation energy,
 (h) the ratio of the radiation energy to the sum of the 
 gas and magnetic energies,
 (i) and the ratio of the gas temperature
 to the radiation temperature.
 All values are time-averaged over $t=6$$-$7 s.
 The white and black arrows in panel (a) indicate
 the velocity vectors whose magnitude exceed the escape velocity.
 The dashed line in panel (b) is the photosphere, 
 at which the optical thickness measured from the upper boundary
 is unity.
 The arrow in panel (g) shows the radiative flux vector.
 \label{fig2DA}   
 }
\end{figure*}
Figure \ref{fig2DA} illustrates
various aspects of the flow structure for model A.
The contours of the density,
the gas temperature,
and the plasma-$\beta$ are shown 
in panels (a), (b), and (c), respectively.
Here, the velocity vectors are overlaid in panel (a) 
with arrows, in which the white arrows indicate
that the velocity exceeds the escape velocity defined as 
$v_{\rm esc}\equiv \left(2GM/R\right)^{1/2}$.
The magnetic energies of the toroidal and poloidal components 
are separately plotted in panels (d) and (e), respectively,
where the poloidal component is given by 
$B_p=\left(B_r^2+B_z^2 \right)^{1/2}$.
The magnetic pitch, $|B_\varphi|/B_p$, 
is shown in panel (f).
Panel (g) indicates the radiation energy,
and the ratio of the radiation energy to 
the sum of the gas and magnetic energies is 
plotted in panel (h). 
Panel (i) shows the ratio of the gas temperature
to the radiation temperature, where
the radiation temperature is calculated as
$T_{\rm rad}=(E_0/a_r)^{1/4}$.
All values are time-averaged over $t=6$$-$7 s,
and both axes are normalized by the Schwarzschild radius.

We find in panel (a) that 
the flow is divided into two regions:
the disk region around the equatorial plane
(characterized by white and red colors)
and the outflow region above the disk region (blue region).
The boundary between the two regions is approximately 
along the line of $z/r\sim 2$.
The matter slowly accretes onto the black hole 
through the disk region with velocity
much less than the free-fall velocity (see small black arrows).
This panel also shows that
the outflow velocity highly exceeds the escape velocity
(see white arrows).
The photosphere, at which the optical thickness measured from 
the upper calculation boundary is about unity, roughly corresponds 
to the green region, i.e., $z/r\sim 2.4$.

We find that the hot, rarefied corona appears above the 
supercritical disk.
Panel (b) indicates that the outflowing matter is very hot
($\gsim 10^9$ K),
whereas the disk consists of relatively hot gas 
($\sim 10^{7-8}$ K).
The gas density is much smaller in the outflow region
than in the disk region (see panel (a)).
Such a hot outflowing corona above the supercritical disk
has also been reported by \citet{O05b},
and is though to affect the spectra 
due to the Comptonization \citep{Kawashima09}.

Panel (c) shows  that the outflow is surrounded by 
the low plasma-$\beta$ region (blue),
in which we find plasma-$\beta$ $<1$.
In such a low plasma-$\beta$ region,
the toroidal component of the magnetic fields,
which seems to be amplified via the differential rotation of gas,
is dominant over the poloidal one (see panels (d) and (e)).
This feature is also clearly demonstrated in panel (f),
in which we see $|B_\varphi|/B_p>10$ around the outflow region.
That is, the outflow is surrounded by the regions 
with strong toroidal magnetic fields.
Inside the outflow,
on the other hand, 
the vertical component of the magnetic fields 
is dominant over the other components. 

Such a magnetic structure is 
quite reminiscent of magnetic-tower jets 
\citep{Lynden94,Lynden96,KMS04}.
However, there is one big difference;
while the magnetic tower jets are accelerated via 
the magnetic-pressure force, 
the outflow in our model A is powered by the radiation force.
This radiatively driven outflow is collimated by the Lorentz force.
Thus, on the basis of global RMHD simulations we proposed 
a novel jet model;
the radiation-pressure driven and magnetically collimated outflow
\citep[see also][]{TOM10}. 

Note that
there must be something to prevent the field from expanding
sideways, thereby realizing the high magnetic energy density 
in the cylindrical region around the black hole. 
In our simulations it is the radiation-pressure force by
geometrically thick flow surrounding the magnetic tower
that confines the magnetic tower.
The magnetic field can thus naturally evolve to form 
such a high concentration from its initial configuration
\citep{TOM10}.

In panel (g), it is found that 
the radiation energy is enhanced in the disk region.
The disk is not only geometrically thick but also optically thick.
Because of numerous scatterings,
the photons generated in the disk 
cannot easily escape from the disk surface.
As a consequence, a large number of photons accumulate in the disk 
region, leading to the high radiation energy density.
The radiation energy is predominant over 
the gas energy,
and also over the magnetic energy in the whole region
(see panel (h)).
The ratio of the radiation energy to 
the sum of the gas and magnetic energies 
is $\gsim 10^3$ in the outflow region
and $\gsim 10$ in the disk region.

Panel (i) indicates that the gas temperature
is nearly equal to the radiation temperature
within the disk (blue).
The absorption opacity is large in the disk region,
since it is sensitive to the density ($\propto \rho^2$).
Thus, the effective gas$-$radiation interaction (emission/absorption)
leads to $T_{\rm gas}\sim T_{\rm rad}$.
In contrast, the gas temperature is much higher than
the radiation temperature in the outflow region,
where the radiative cooling is not efficient
because of the small absorption opacity (small density).
Thus, the gas temperature is kept high
(see the next subsection for more quantitative descriptions).

\subsubsection{Vertical Structure}
\label{VerA}
\begin{figure}
 \epsscale{1.18}
 \plotone{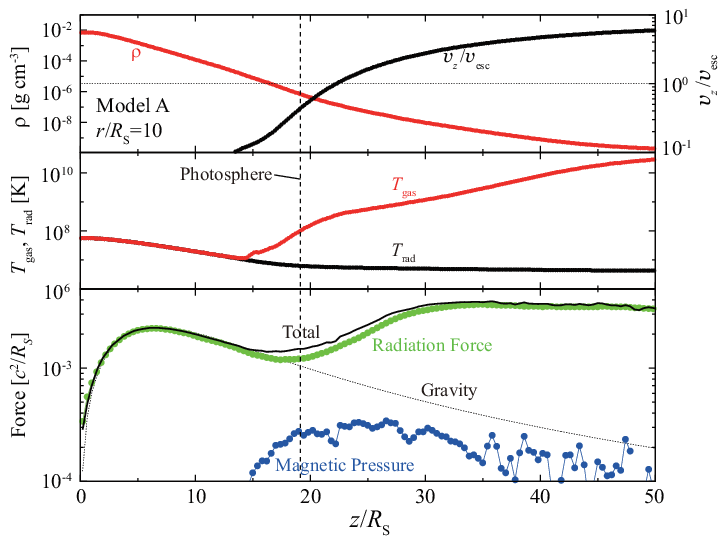}
 \caption{
 \label{figverA}   
 Vertical structure at $r=10R_{\rm S}$ for model A.
 Top panel:
 the density (red) and vertical component of the velocity normalized 
 by the escape velocity (black).
 Middle panel:
 the gas and radiation temperatures.
 Bottom panel: the vertical components of the 
 gravity (dotted line), 
 the radiation force (green filled circle),
 and the magnetic-pressure force (blue filled circle).
 The sum of the vertical forces by the radiation,
 the magnetic pressure, the magnetic tension, 
 and the gas pressure is plotted by the black solid line (total).
 Here, the gas-pressure force (red filled circle) 
 and the magnetic-tension force (blue open circle)
 are not represented, since they are so small or negative.
 All values are time-averaged over 6$-$7 s.
 }
\end{figure}
In the top panel of Figure \ref{figverA},
we plot the vertical profiles of the density (red)
and the vertical component of the velocity (black)
for $r/R_{\rm S}=10$.
Here, they are time-averaged over $t=5$$-$7.5 s.
We find that 
the density decreases with an increase of $z$.
The upward velocity increases as $z$ increases
and it exceeds the escape velocity at $z\gsim 20R_{\rm S}$.
The sonic point is around $z=20R_{\rm S}$.
While 
the gas temperature is comparable to the radiation temperature
due to the effective gas$-$radiation interaction at $z\lsim 15R_{\rm S}$,
we find $T_{\rm gas} \gg T_{\rm rad}$ at the upper region
of $z\gsim 20R_{\rm S}$,
since radiative cooling is inefficient 
because of smaller density (emissivity) (see the middle panel).

As we have mentioned in Section \ref{2DA},
the supercritical disk ejects
high-velocity outflows driven by the radiation force.
This is clearly shown in the bottom panel
of Figure \ref{figverA},
in which we plot vertical components of 
the gravity (dotted line), the radiation force 
(green circle), the magnetic-pressure force (blue circle).
The solid black line (total) in the bottom panel
indicates the sum of 
the vertical forces by the radiation,
the magnetic pressure, the magnetic tension, 
and the gas pressure.
(The gas-pressure force and the magnetic-tension force
are so small that they do not appear in this figure.)
While the radiation force is nearly balanced with the gravity
at $z\lsim 20R_{\rm S}$ (disk region),
it largely exceeds the gravity in the 
upper regions at $z\gsim 20R_{\rm S}$,
producing the high-velocity outflows.
The radiative flux is roughly estimated 
to be $cE_0$ in the optically thin, 
outflow region above the disk,
whereas it is reduced to be $cE_0/\tau$ 
in the optically thick disk region.
Hence, the radiation force suddenly 
increases above
around the disk surface, $z\sim 20R_{\rm S}$,
although the radiation energy density itself ($E_0$) is smaller
in the outflow region than in the disk region 
(see panel (g) of Figure \ref{fig2DA}).
In the case of the supercritical flows,
the radiation force supports the disks
and accelerates the outflows.
The forces via the magnetic pressure,
the magnetic tension, and the gas pressure is
too small to influence the acceleration of the outflow.
However, the magnetic pressure and the magnetic tension
(Lorentz force) works 
in a direction parallel to the disk plane and thus
collimates the outflow
\citep[see][]{TOM10}.

\subsubsection{Dissipation Rate}
\label{dissA}
\begin{figure}
 \epsscale{1.1}
 \plotone{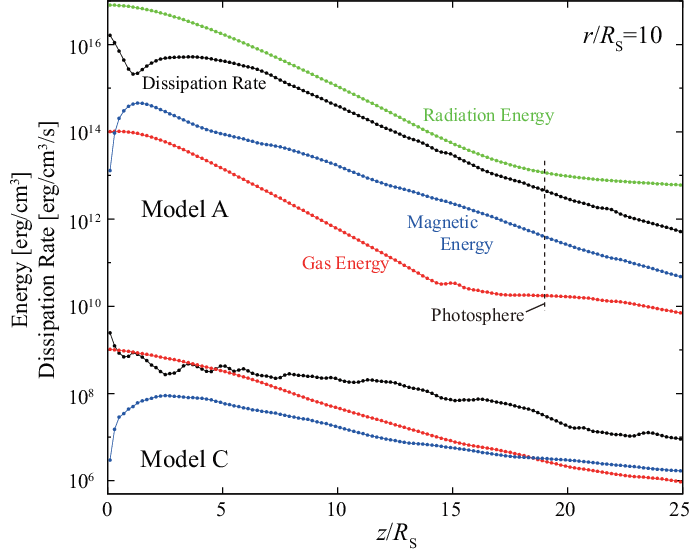}
 \caption{
 Vertical profiles of 
 the energy dissipation rate (black)
 the gas (red), radiation (green), and magnetic (blue) energies
 at $r=10R_{\rm S}$.
 They are time-averaged over 5$-$7.5 s.
 The radiation energy in model C is not plotted,
 since it is too small.
 }
 \label{diss}
\end{figure}
In Figure \ref{diss},
we plot the vertical profiles of the dissipation rate,
$4\pi\eta J^2/c^2$ (black),
as well as those of the gas (red), radiation (green), 
and magnetic energies (blue) for model A 
(see upper four lines).
They are time-averaged over 5$-$7.5 sec.
All the values tend to decrease with an increase of $z$.
In the region of $z\lsim 20R_{\rm S}$,
we find that the gradient of the dissipation rate 
is smaller than that of the gas energy
and that of the radiation energy.
It implies that the traditional $\alpha$-viscosity model 
does not precisely describe the vertical structure of the disk,
since it states that the dissipation rate is proportional to 
the gas (or radiation) pressure.
We find that 
the profile of the dissipation rate is roughly on parallel with
that of the magnetic energy
\citep[see also][for the cases of local RMHD simulations]
{Turner03,Hirose06}. 

The radiation energy shows a rather flat distribution above 
$z\gsim 20R_{\rm S}$.
Since the radiation energy is transported via the diffusion
in the disk region, 
while the photons freely go out above the disk,
the slope of the profile of the radiation energy changes 
across the disk surface at $z\sim 20R_{\rm S}$.

In figure \ref{diss}, we also find that,
while the dissipation rate is enhanced near the equatorial plane
in both models,
the magnetic energy suddenly decreases toward $z=0$.
However cautions should be taken here, 
since this might be caused by the particular boundary conditions
with respect to the equatorial plane, where 
we require that $B_r$ and $B_\varphi$ are antisymmetric. 
Such a condition enhances the reconnection,
leading to the decrease of the magnetic energy
and the increase of the dissipation rate.
In fact, \citet{Hirose06} showed 
that the magnetic energy and the dissipation rate 
are nearly constant across the equatorial plane
by simulations without 
employing the equatorial-plane symmetries.

Although the traditional $\alpha$-viscosity model 
does not precisely describe the vertical dissipation profile,
the $r$$-$$\varphi$ component of the shear-stress tensor
at around the equatorial plane is roughly proportional to the pressure
when we examine the time variations of these two quantities.
In the previous paper \citep{O09},
we investigated the time evolution of the magnetic torque
of $z\lsim 3R_{\rm S}$, and demonstrated that
the torque is roughly proportional 
to the total pressure but with some scatters.

\subsubsection{Gaussian or Polytropic?}
\label{GaussA}
\begin{figure}
 \epsscale{1.}
 \plotone{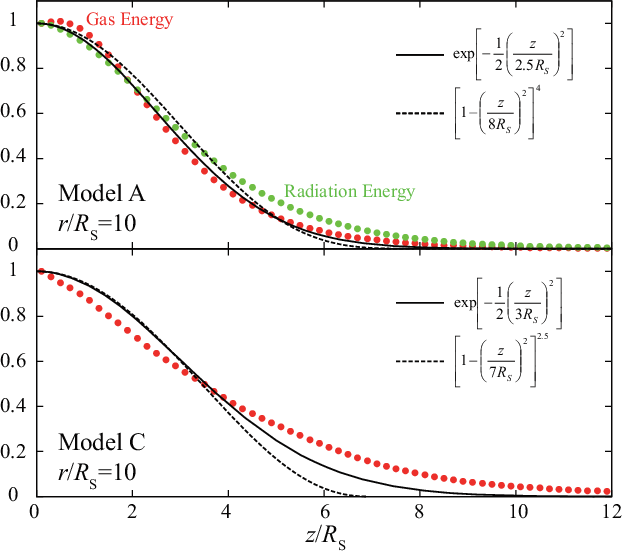}
 \caption{
 Gas and radiation energies of model A (top) and 
 gas energy of model C (bottom)
 normalized by the each value at $z=0$.
 Solid and dotted lines indicate 
 the Gaussian and polytropic relations,
respectively.
 }
 \label{fit}
\end{figure}
Finally, we compare the simulated vertical structure
with that obtained by simple analytic models on the
assumption of the hydrostatic balance.
If the disks are vertically isothermal,
the energy (pressure) profile is given by 
a Gaussian profile of 
$\exp\left(-z^2/2H^2 \right)$ with 
$H$ being the disk half-thickness.
Such a function with $H=2.5R_{\rm S}$ 
roughly reproduces the energy distributions of the 
flow in model A, as is seen 
in the top panel of Figure \ref{fit}, which represents
gas and radiation energies of model A 
normalized by the values at $z=0$.
If we use a polytropic relation 
($p \propto \rho^{(N+1)/N}$ with $N$ being the polytropic index)
instead of the isothermal assumption, on the other hand,
the vertical hydrostatic balance leads
that the energy is proportional to a function of 
$\left(1-z^2/2H^2 \right)^{N+1}$.
Such a function with $H=8R_{\rm S}$ can also 
give good fits to the energy profiles of model A
(see the upper panel in Figure \ref{fit}).
Here, we employ $N=3$ for model A 
since the disk is radiation-pressure-dominated.
We note that the disk half-thickness at $r=10R_{\rm S}$ is 
$H\sim \left(c_{\rm S}/v_{\rm Kep}\right) r\sim 5R_{\rm S}$
and are in rough coincidence with the above values,
where $c_{\rm S}$ is the sound velocity at $(r, z)=(10R_{\rm S}, 0)$.

\subsection{Model B}
\label{modelB}
\subsubsection{Overview}
\label{overB}
As is already mentioned, 
accretion flows at $r\gsim 7R_{\rm S}$
in our model B ($\rho_0=10^{-4}\, \rm g\, cm^{-3}$)
resemble the standard disk.
The efficient radiative cooling leads to the formation of a 
geometrically thin 
but optically thick disk (see Figure \ref{figview}).
The disk in model B is moderately dense
as shown in the top panel of Figure \ref{figmid}.
The gas temperature of $\sim 10^6$K at $r\gsim 7 R_{\rm S}$
is close to the value predicted by the standard disk model
(see the middle panel of Figure \ref{figmid}).
However, the disk is truncated 
around $r\sim 7 R_{\rm S}$ (which we call the truncation radius).
Because of small density of the accretion flow
near the black hole,
the emissivity is greatly reduced,
which is responsible for 
a decoupling of the gas and radiation temperatures,
$T_{\rm gas}\gg T_{\rm rad}$ at $r\lsim 5 R_{\rm S}$.
The bottom panel shows that 
the inflow velocity ($-v_r$) increases
as the flow approaches the black hole.
The rotation velocity 
is very close to the Keplerian velocity
(within 10\%),
although we do not represent it in Figure \ref{figmid}.

We find in Figure \ref{figML}
that the photon luminosity is well below the Eddington luminosity,
$\sim 2 \times 10^{-4} L_{\rm E}$,
for the mass accretion rate is $\sim 5\times 10^{-3} L_{\rm E}/c^2$.
In this model, the photon trapping effect is negligible.
The energy conversion efficiency,
$\eta \equiv L_{\rm ph}/\dot{M}_{\rm acc} c^2 \sim 0.04$,
is the largest in three models.
This is also consistent with the prediction of 
one-dimensional accretion disk study, whereby
the radiative cooling is more effective 
in the standard disk model than the slim disk model and RIAF.
However, the simulated value of $\eta$ is smaller than 
the prediction of the standard disk model, $\sim 0.1$.
This result might be caused by the disk truncation
around $r\sim 7R_{\rm S}$,
within which the density is too small 
for emission to be efficient (will be discussed in Section \ref{SED}).
If we were to employ higher density normalization,
both the photon luminosity 
and the conversion efficiency might increase,
since then the truncation radius would decrease or even disappear.
Such a disk would be more consistent with the standard disk model.

Whereas the disk wind is not predicted by
the standard disk model,
a significant amount of matter is blown away in our model B.
As we have already mentioned in Section \ref{overview},
the matter above the disk rotates and goes upward.
In other words, outflow material possesses
substantial angular momentum, taking it away from the underlying disk.
Figure \ref{figML}
illustrates that time-averaged mass-outflow rate
is about $10^{-4} L_{\rm E}/c^2$ (see the top panel),
which is a few \% of the mass accretion rate
(see the bottom panel).
We also find in the bottom panel
that $L_{\rm kin}$ is a few \% of $L_{\rm ph}$.
As is the case in model A,
the disk in model B releases energy via radiation rather than via outflows.
Here we note a possibility that 
the outflowing matter with high velocity
might originate not from the outer cold part of the disk
($r\gsim 7R_{\rm S}$)
but from the inner hot part of the disk ($r\lsim 5R_{\rm S}$).
The matter ejected from the outer cold part of the disk 
might produce low-velocity outflow with a wide opening angle.
At this moment it is technically very difficult to identify
the launching points of the high- and low-velocity outflows.
Note that there is no clear evidence for the collimation of
the outflows in model B.

\subsubsection{Two-dimensional Structure}
\begin{figure*}
 \epsscale{0.825}
 \plotone{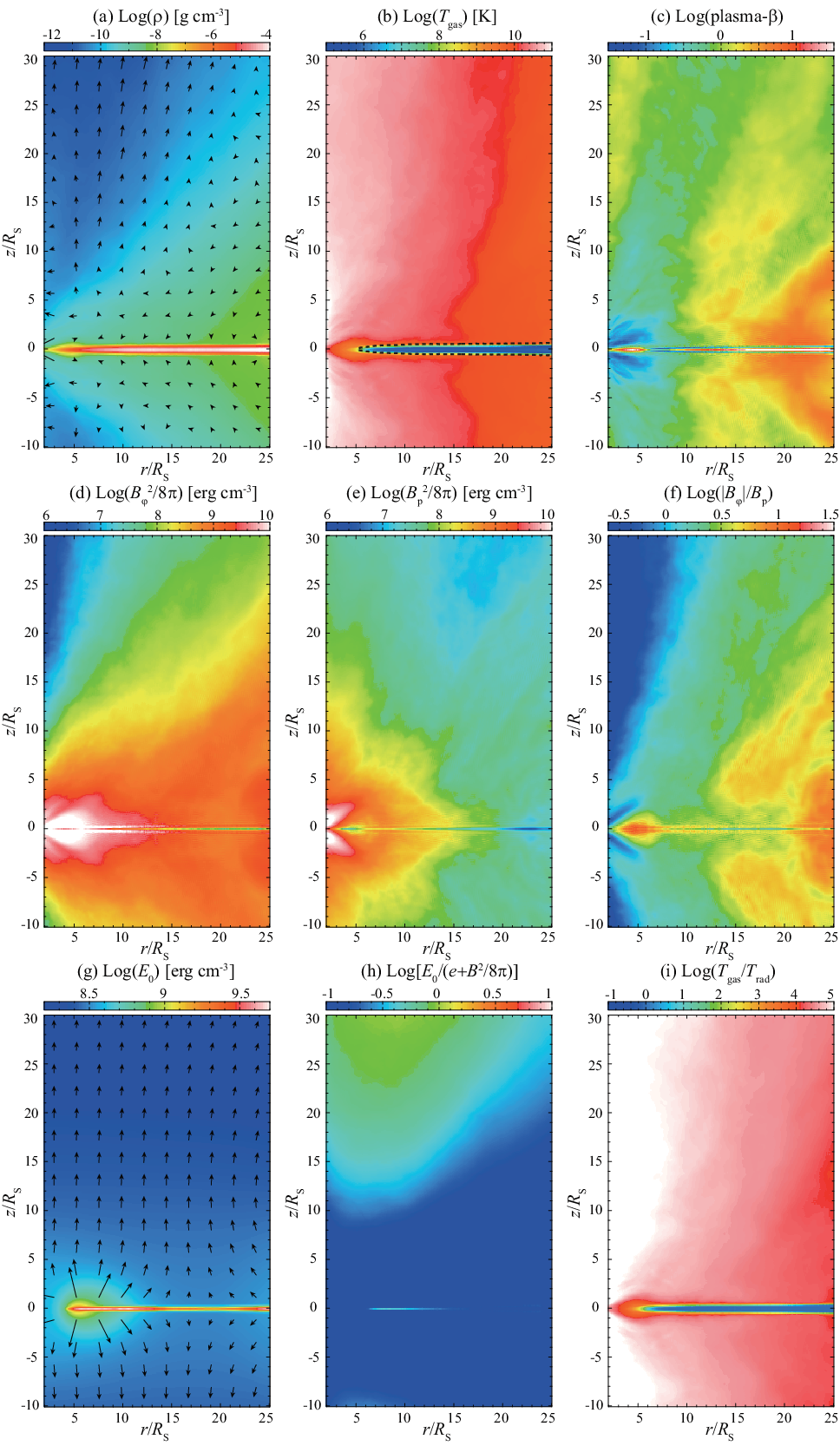}
 \caption{
 Same as Figure \ref{fig2DA} but for model B.
 Each quantity is time-averaged over $t=9$$-$10 s.
 Note the different plot area from that of Figure \ref{fig2DA}.
 \label{fig2DB}   
 }
\end{figure*}

%
Two-dimensional structure of the flow in model B
is presented in Figure \ref{fig2DB}.
Each physical quantity is
time-averaged over $t=9$$-$10 s.
Panel (a) 
clearly shows that a geometrically thin disk is located
at around the equatorial plane 
(indicated by the white and red colors).
The disk thickness is very small and is everywhere less than $\sim 1 R_{\rm S}$.
The gas of the disk effectively cools by emitting radiation,
as shown in panel (b),
except at $r \lsim 5R_{\rm S}$.
The gas temperature is very high above and below the disk,
since the low density leads to a low cooling rate 
(emissivity).
Such high-temperature and low-density matter forms
a corona surrounding the cold disk and Compton upscatters
seed photons generated within scale height the disk.

Panel (c) shows that
a relatively low plasma-$\beta$ region appears 
along the line of $|z|/r\sim 2$
(green and light blue).
We find that 
the matter goes upward in this region 
(see vectors in the blue region in panel (a)).
Although the time-averaged velocities are less than
the escape velocity,
the upward velocity intermittently exceeds the
escape velocity. 
As a result, the time-averaged mass-outflow rate is 
$\dot{M}_{\rm out} \sim 10^{-4} L_{\rm E}/c^2$
(see Figure \ref{figML}).
Panel (f) shows that 
this outflow is surrounded by the regions of $|B_\varphi|/B_{\rm p}>1$,
while the poloidal (vertical) component of the magnetic fields 
is dominant at the 
very vicinity of the rotation axis.
These features can be seen in panels (d) and (e).
The toroidal field is enhanced mainly 
near the disk, $|z|\lsim 10R_{\rm S}$ (see panel (d)).
The poloidal component, in contrast, tends to be enhanced
near the black hole (see panel (e)).
Again, something like magnetic tower structure appears
in model B, as well.
Here we note that the higher plasma-$\beta$ 
and the larger magnetic pitch 
at the outer region ($r\gsim 20R_{\rm S}$ and $|z| \lsim 15R_{\rm S}$) 
are thought to be under the influence of the initial torus.

We find in panel (g) that
the radiation energy is enhanced within the disk (white and red),
since the disk is optically thick except at $r\lsim 7R_{\rm S}$.
The optical thickness of the disk atmosphere 
is, by definition, smaller than unity,
implying that the photons emitted at the disk surface
can freely go out.
Thus, the radiation energy is small above and below the disk (blue).
In contrast with model A, 
the radiation energy is smaller than the gas energy and
the magnetic energy (see panel (h)).

Panel (i) shows $T_{\rm gas} \sim T_{\rm rad}$
in the disk at $r\gsim 7R_{\rm S}$ (blue),
and $T_{\rm gas} \gg T_{\rm rad}$
in the other regions (white and red).
This is because the gas$-$radiation interaction is effective only in the dense
region as we have already mentioned.
Such a feature is similar to that in model A,
though the disk is geometrically thin in model B.

\subsubsection{Vertical Structure}
\begin{figure}
 \epsscale{1.18}
 \plotone{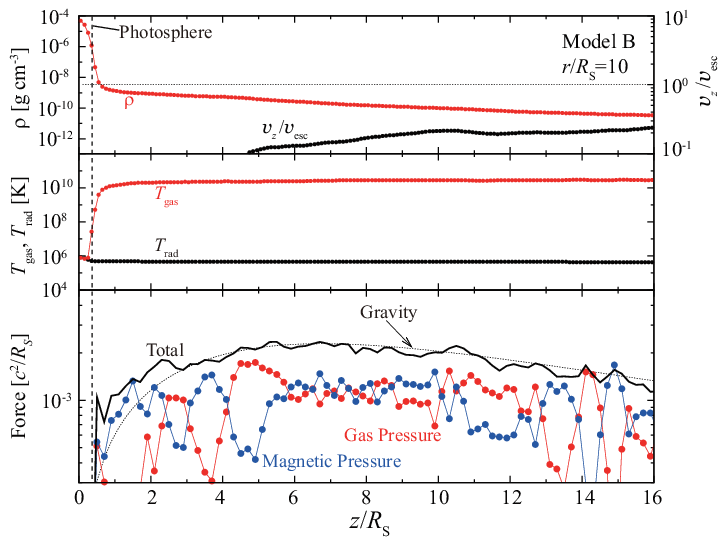}
 \caption{
 Same as Figure \ref{figverA} but for model B.
 Each quantity is time-averaged over $t=9$$-$10 s.
 Here, the radiation force (green filled circle) 
 and the magnetic-tension force (blue open circle)
 are not represented, since they are so small or negative.
 Here we note that the disk region contains 5$-$6 grid points 
 (excluding the point at the boundary). 
 Although this number is not big, it can marginally resolve 
 the vertical structure.
 \label{figverB}   
 }
\end{figure}
Figure \ref{figverB} is the same as Figure \ref{figverA}
but for model B and $t=7.5$$-$10 s.
In the bottom panel, 
the red circles indicate the gas-pressure force.
In the top panel,
the high-density regions of $z<R_{\rm S}$
correspond to the thin disk as we have shown in Figure \ref{fig2DB}.
The density slowly decreases with an increase of $z$ 
above the disk.
This is because 
the gas temperature is very high 
and so is the scale height in the low-density region at $z>R_{\rm S}$
(middle panel).
We find $T_{\rm gas} \sim T_{\rm rad}$ in the disk region.

We find that sum of the gas and magnetic-pressure forces roughly 
balance with the gravity at $z\gsim 3R_{\rm S}$ (see the bottom panel).
The time-averaged upward velocity
is much smaller than the escape velocity 
in this region (see the top panel).
Note, however, that the high-velocity outflows occasionally appear.

How is the matter ejected from the disk?
It is mainly by the magnetic-pressure force.
The bottom panel shows that 
the hydrostatic balance breaks down around the disk surface,
$z=1-3R_{\rm S}$,
where the magnetic-pressure force exceeds the gravity,
leading to the mass ejection from the disk surface.
In contrast with model A, 
the radiation force is negligible.

Although we used small grid spacings, $\Delta r=\Delta z=0.1R_{\rm S}$,
in our simulations for model B,
we note that the simulations of even higher resolution are required
to investigate the detailed structure of the geometrically thin disk.
Such calculations will be performed in a future study.

\subsection{Model C}
\label{modelC}
\subsubsection{Overview}
\label{overC}
Flows in model C ($\rho_0=10^{-8}\, \rm g\, cm^{-3}$)
correspond to the RIAF.
The geometrically thick disk is supported by gas pressure 
in cooperation with the magnetic pressure
(see Sections \ref{2DC} and \ref{VerC}).
As shown in Figure \ref{figmid},
the flow is entirely occupied by 
very hot and rarefied plasma.
Since the density is too low for radiative cooling to be effective,
a decoupling of the gas and radiation temperatures,
$T_{\rm gas} \gg T_{\rm rad}$, 
occurs in the whole region.
The bottom panel in Figure \ref{figmid} shows that 
the accretion velocity is relatively high at $r\lsim 10R_{\rm S}$
in comparison with those of the flows in models A and B 

Since the optical thickness is very small ($\tau \sim 10^{-4}$), 
the flow in model C is radiatively inefficient and faint.
The photon luminosity is much smaller than the Eddington luminosity,
$L_{\rm ph} \ll 10^{-8} L_{\rm E}$ (see Figure \ref{figML}).
The energy conversion efficiency in model C 
($\eta \sim 10^{-5}$) 
is much smaller than those in models A and B.

The helical streamlines 
seen in Figure \ref{figview} indicate
the formation of jets around the rotation axis.
The mass-outflow rate is 
about $10$ \% of the mass accretion rate
(see Figure \ref{figML}).
Interestingly, we find
$L_{\rm kin} > 10^2 L_{\rm ph}$ in model C
(cf. $L_{\rm kin}<0.1L_{\rm ph}$ in models A and B).
Hence, the radiatively inefficient disk 
($\dot{M}_{\rm acc}\ll L_{\rm E}/c^2$)
loses the energy more via jets rather than via radiation.

\subsubsection{Two-dimensional Structure}
\label{2DC}
\begin{figure*}
 \epsscale{0.825}
 \plotone{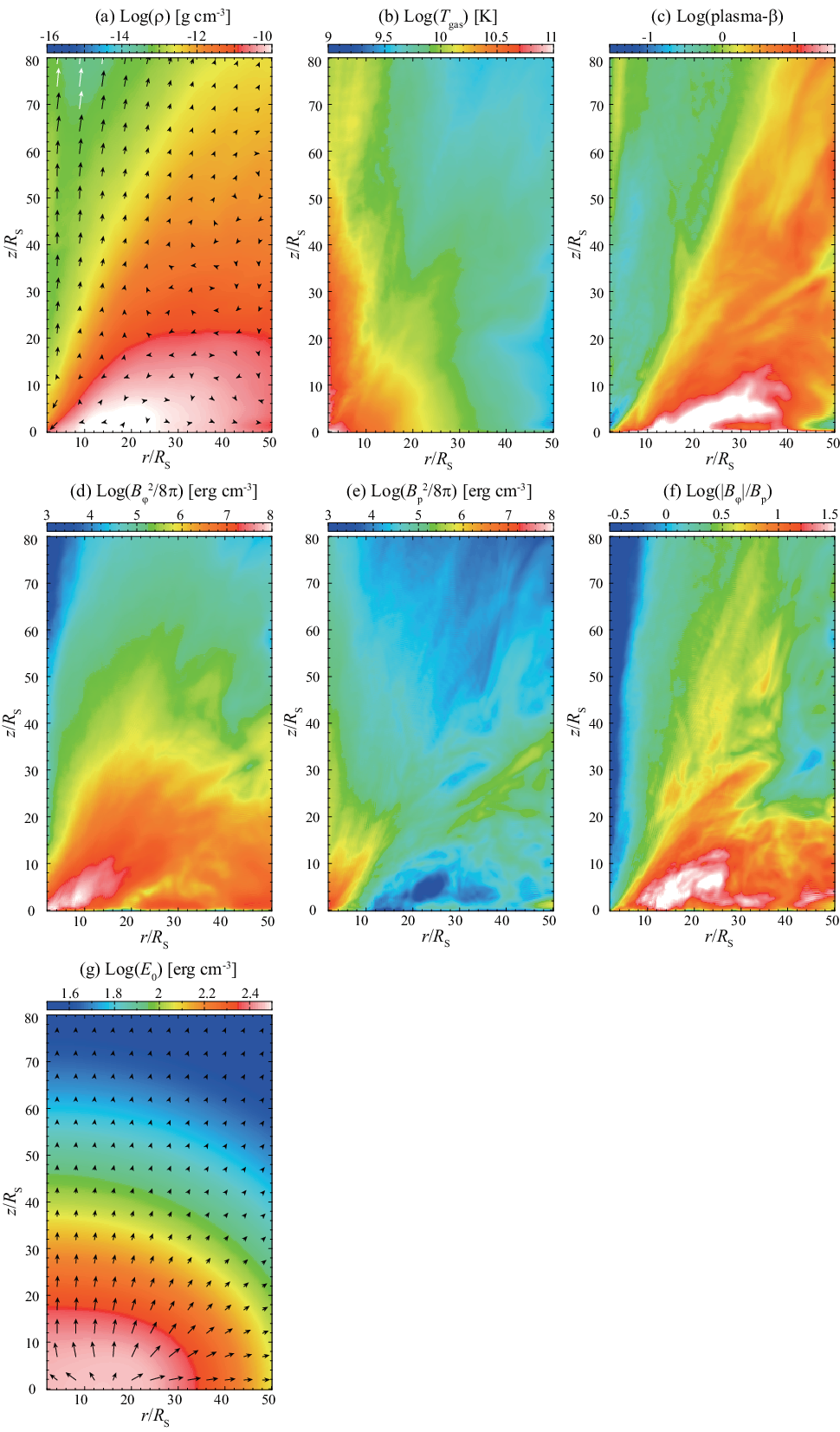}
 \caption{
 Same as panels (a)$-$(f) of Figure \ref{fig2DA}, but for model C.
 Here, note that there is no photosphere, 
 since the flow is optically thin.
 \label{fig2DC}   
 }
\end{figure*}
Figure \ref{fig2DC} shows the two-dimensional structure
of the flow in model C.
This figure is the same as panels (a)$-$(f) 
of Figure \ref{fig2DA} but for model C.
The white and red colors in panel (a) indicate
a geometrically thick disk (disk region).
The entire zone can be divided to the disk and outflow regions
by the line of $z/r \sim 2$.
The gas temperature highly exceeds $10^9$ K in the whole region
(panel (b)).
Here we note that we might overestimate 
the gas temperature, since the cooling via the Compton scattering
is not taken into consideration.
Also, the electron temperature might deviate from the ion temperatures,
although we treat the matter as a one-temperature plasma.
The disk is mainly supported by the gas pressure.
The gas energy is dominant over the magnetic pressure
in the disk region (see panel (c)).

In panel (a), we also find 
that the matter with relatively small density 
is blown away above the disk, i.e., $z/r\gsim 2$. 
(The outflow region is characterized by the green and yellow colors.)
The time-averaged outward velocity
exceeds the escape velocity only in the region
of $z\gsim 70R_{\rm S}$ in panel (a) (white vectors).
Note, however, that this figure only shows
the time average, while the simulations exhibit significant
time variations in $v_R$. 
In fact, we find that 
the high-velocity outflows ($v_R>v_{\rm esc}$) 
occasionally appear at $z\gsim 10R_{\rm S}$ near the rotation axis.
The outflow is mainly accelerated via
the magnetic pressure, 
in cooperation with the gas pressure
(we will discuss below).
The radiation force is negligible.

Panel (c) indicates plasma-$\beta \lsim 1$ 
in the outflow region,
whereas $\beta>1$ in the disk region.
The toroidal magnetic fields are mainly amplified in the disk region,
while poloidal fields strengthen around the rotation axis
(see panels (d) and (e)).
Hence, the magnetic pitch, $|B_\varphi|/B_{\rm p}$, 
tends to increase as an increase of $r$.
Panel (f) shows that the pitch becomes large
around the outer edge of the outflow region
located long the line of $z/r \sim 2$
(see the yellow area).
This implies that the magnetic field lines are strongly
coiled around the spine of the outflow;
that is, helical magnetic fields form.
Thus, the outflow in our model C seems to be 
magnetic-tower jets, which have been reported by 
the three-dimensional MHD simulations by 
\citeauthor{KMS04}
(\citeyear{KMS04}; see also
\citeauthor{Lynden94} \citeyear{Lynden94};
\citeauthor{Lynden96} \citeyear{Lynden96},
for the original proposal).

Since the disk is optically thin,
the photons can freely escape from the disk,
producing a quasi-spherical distribution of 
the radiation energy density,
as is clearly shown in panel (g).
Again, 
the radiation force is negligible
due to the small radiation energy and small opacity.

\subsubsection{Vertical Structure}
\label{VerC}
\begin{figure}
 \epsscale{1.18}
 \plotone{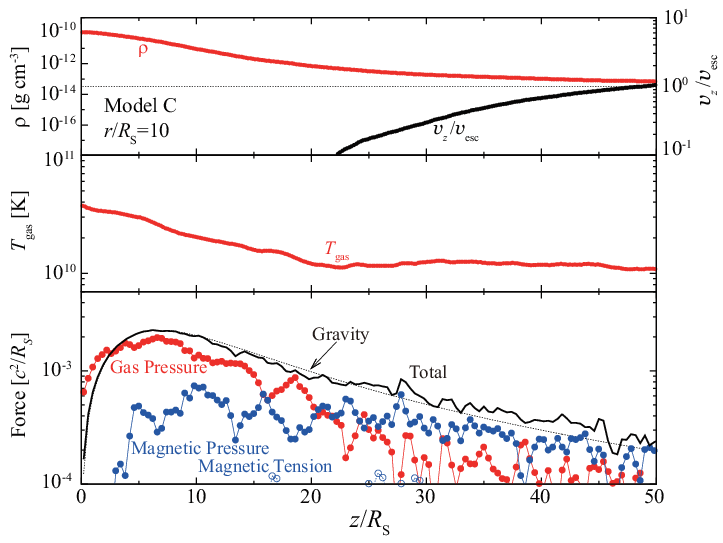}
 \caption{
 Same as Figure \ref{figverA} but for model C.
 Here, the radiation force (green filled circle) 
 is not shown, since they are negligibly small. 
 Note that the gas pressure exceeds the total pressure at
 small height, $z\leq 2R_{\rm S}$ (see the bottom panel).
 This is because the vertical force by the magnetic pressure is negative there.
 \label{figverC}   
 }
\end{figure}
Figure \ref{figverC} is the same as Figure \ref{figverA}
but for model C.
The density and the gas temperature monotonically
decrease with an increase of $z$.
The open blue circles indicate the vertical component
of the magnetic-tension force.
We find that the geometrically thick disk at $z\lsim 20R_{\rm S}$ 
is mainly supported by the gas pressure,
achieving the hydrostatic balance
(see the bottom panel).
The magnetic-pressure force 
is dominant over the gas-pressure force
above the disk, $z\gsim 20R_{\rm S}$.
In this region, the total force (black line) exceeds the gravity,
implying that the matter is accelerated upward.
That is, the magnetic-pressure force 
in cooperation with the gas-pressure force
drives the outflows above and below the disk. 
The outflow is thus accelerated upward;
its velocity exceeds the sound velocity at $z\sim 30R_{\rm S}$
and reaches the escape velocity at $z\sim 50R_{\rm S}$.
Both the radiation force and the 
magnetic-tension force are found to be negligible.

Similarly to model A (see Figure \ref{figverA}),  
we find in model C that the disk is in hydrostatic balance and
the matter is blown away from the disk surface.
However, the radiatively driven outflows
in model A is more powerful than
the magnetically driven outflows in model C
(The kinetic luminosity, $L_{\rm kin}$, is larger in model A
than in model C).
While the radiation force largely exceeds
the gravity in model A,
the upward force only slightly exceeds the gravitational force
in model C (see the bottom panel in Figure \ref{figverC}).
As a result,
low-luminosity disks with 
$\dot{M}_{\rm acc}\ll L_{\rm E}/c^2$
cannot produce powerful high-velocity outflows 
in a steady fashion, but only occasionally.

\subsubsection{Dissipation Rate}
\label{dissC}
The three lower lines in Figure \ref{diss} represent
the vertical profiles of the dissipation rate (black), 
gas (red) and magnetic energies (blue) for model C,
which are time-averaged over 5$-$7.5 s.
Here, we note that the radiation energy 
is never appreciable in model C,
so it is not plotted.
We find that
the slope of the profile of the dissipation rate is 
nearly equal to that of the magnetic energy,
and flatter than that of the gas energy.
Such a feature
implies that the traditional $\alpha$-viscosity model 
does not precisely describe the vertical flow structure,
as was already noticed in model A.
As we have discussed in Section \ref{dissA},
the boundary condition
with respect to the equatorial plane
seems responsible for the enhanced dissipation rate,
leading to the drop of the magnetic energy at $z\sim 0$.

\subsubsection{Gaussian or Polytropic?}
\label{GaussC}
In Section \ref{GaussA},
we have mentioned that 
the energy distributions of the flow in model A
agrees with the profiles of the disks in hydrostatic balance
(isothermal and polytropic).
However, we cannot fit the profile of the gas energy of model C
by the Gaussian profile (with $H=3R_{\rm S}$)
nor by the polytropic relation (with $H=7R_{\rm S}$,
see the bottom panel in Figure \ref{fit}).
Here, $N=1.5$ is used because the gas pressure is predominant
for the disk of model C.
Even if we change $H$, fitting is not successful.
The drop of the gas energy is more rapid at small $z$ ($\lsim 4R_{\rm S}$) 
and steeper above.
Such deviations may be caused by the particular magnetic energy profile
which is not simply proportional to gas pressure.
Finer-mesh calculations are necessary to confirm
if this is really the case.


\section{Discussions}
\label{discussions}
\subsection{Comparison with Observations: Outflow}
Our simulations demonstrate that
powerful outflows appear
in the super- or near-critical flows (model A).
This result seems to be consistent with the observations of 
bright black hole objects with high Eddington ratio, 
such as narrow-line Seyfert 1 galaxies (NLS1s)
and microquasars.
The NLS1s, which are thought to contain
relatively less massive black holes and thus 
to be high $L_{\rm ph}/L_{\rm E}$ systems,
are usually radio-quiet. 
However, some of them have been reported to be radio-loud.
\citet{Doi06}, for example, concluded by 
very long baseline interferometry observations that 
a radio-loud NLS1, J094857.3+002225, possess relativistic jets
\citep[see also][]{Zhou03}. 
By contrast, there are plenty of such sources in our Galaxy.
Microquasars, such as GRS 1915+105, also show 
relativistic jets in high-luminosity state
\citep{Fender04} and 
SS433 is another good example for 
supercritical accretion flow with powerful jets.

Our model A may explain 
the nebulae (or bubbles) around ultraluminous X-ray sources (ULXs) 
or microquasars.
\citet{Pakull10} proposed by 
X-ray/Optical observations that 
the large nebula, S26, is produced by energy injection via powerful jets,
of which the kinetic luminosity 
is estimated to exceed the photon luminosity, 
$L_{\rm kin}>L_{\rm ph}$,
by three orders of magnitude
\citep[see also][]{Pakull08}.
The supercritical flows will be observed 
as low luminosity objects ($L_{\rm ph}<L_{\rm kin}$)
near the edge-on view,
since the emission is mildly collimated 
and since the radiation from the innermost region can
be obscured by the outer disk.
If this is the case, 
the large $L_{\rm kin}/L_{\rm ph}$ ratio of S26 can be 
explained by our model A.
By solving the radiation transfer equation,
\citet{O05b} demonstrated that the apparent luminosity in the edge-on view 
could be more than 10 times smaller than that in the face-on view.
In this calculation, large luminosity comes from the middle part of 
the disk at $r\gsim 100R_{\rm S}$
by the limitation of the computational box.
If we can make simulations with a much more extended computational box,
an even smaller value of $L_{\rm ph}$ is expected
at the edge-on view.
Since $L_{\rm kin}$ is on the order of $L_{\rm E}$ regardless of 
the viewing angle, 
the large $L_{\rm kin}/L_{\rm ph}$ is, hence, expected.


Our model B shows the disk wind and the geometrically thin disk,
which was not predicted 
in the framework of the standard disk model.
The recent X-ray observations of black hole binaries (BHBs)
reveal the presence of the blueshifted absorption lines,
meaning the occurrence of the disk wind 
from the geometrically thin disk.
\citet{Miller06a,Miller06b} concluded by the observations of 
GRO J1655-40
and a black hole candidate, H1743-322,
that the matter with $\rho\sim 10^{-9}\, \rm g\, cm^{-3}$
ejected from the geometrically thin disk
with the speed of $500-1600\, \rm km\,s^{-1}$.
The outflow velocity of 
the X-Ray transient, 4U 1630-472, 
has been also reported to be $\sim 10^3\, \rm km\,s^{-1}$
\citep{Kubota07}.
The present study can account for such observations.
Our model B shows that 
the gas with $\rho\sim 10^{-9}\, \rm g\, cm^{-3}$
goes out at the speed of $\sim 10^3\, \rm km\,s^{-1}$
(see Figures \ref{fig2DB} and \ref{figverB}).

We find the magnetic-tower jet in model C.
In contrast with models A and B in which
we find $L_{\rm kin} < L_{\rm ph}$,
the kinetic luminosity largely exceeds the photon luminosity
in model C, $L_{\rm kin} \gg L_{\rm ph}$
(see Figure \ref{figML}).
That is, the radiatively inefficient flow
releases the energy mainly via the jet.
This result can be understood by the predictions,
whereby the kinetic luminosity is proportional to the 
mass accretion rate, 
$L_{\rm kin} \propto \dot{M}_{\rm acc}$,
but $L_{\rm ph} \propto \dot{M}_{\rm acc}^2$
in the radiatively inefficient regime
\citep[][see also Figure \ref{figML}]{Fender04, NY05, KFM08}.

\subsection{Comparison with Observations: Spectra}
\label{SED}
Our model A might resolve 
the outstanding issue regarding the central engine of ULXs.
As for their central engine,
two competing models have been discussed over the past decade:
subcritical accretion onto intermediate-mass black holes
with the black hole mass greatly exceeding $100 M_\odot$
\citep[e.g.,][]{Max00,Miller04}
and supercritical accretion onto stellar-mass black holes
with mass of $3-20 M_\odot$
\citep[e.g.,][]{King01, Wat+01}.
In addition, there is the possibility that some ULXs may be 
powered by stellar (but not stellar-mass) black holes in the 
$30-80 M_{\odot}$ range, 
formed in low metallicity environments, with near-critical or
slightly supercritical accretion flows
\citep[e.g.,][]{Z+09,M+,B+10}.
It is thus crucially important to show theoretically
how much (apparent) luminosities can be achieved
by supercritical accretion.

The photon luminosity in model A is calculated 
based on the radiative flux mildly collimated within
the polar angle of $\theta=\tan^{-1}(25R_{\rm S}/60R_{\rm S})\sim 23^\circ$
as we have mentioned in Section \ref{overA}.
Hence, the flows in model A would be identified 
as extremely luminous objects of 
$\sim 13L_{\rm ph}=22L_{\rm E}$
for the accretion rate of $\sim 300 L_{\rm E}/c^2$,
or $1.5\times 10^{41}\, \rm erg\, s^{-1}$,
for the case of a black hole of $50M_\odot$,
in the face-on view,
if an observer assumes the isotropic radiation field.
Even higher apparent (isotropic) luminosities are
feasible for higher accretion rates,
since there is practically no limit to the
mass accretion rate \citep{O06}.

The outflowing matter in model A is very hot, 
$T_{\rm gas} \gsim 10^9$ K,
and dense, $\tau_{\rm es}\sim 1$
with $\tau_{\rm es}$ being the Thomson scattering optical depth,
while the disk is relatively cold, $T_{\rm gas} \sim 10^{7-8}$ K
(see Figures \ref{fig2DA} and \ref{figverA}).
Thus, the hot outflowing matter would Compton upscatter photons 
from the disk surface. 
\citet{Kawashima09} have shown that 
at high luminosities, $L_{\rm ph} \gsim L_{\rm E}$,
a very hard spectral state appeared  
due to the effective inverse Compton scattering 
by the radiatively driven outflow.
Their result can explain the hard X-ray spectra
of ULXs \citep[e.g.,][]{Berghea08}
and seems to correspond to the ultraluminous state of ULXs
\citep{Gladstone09}.
In addition,
since an inner part of the disk is obscured by the strong outflow,
the hot thermal component might be invisible.
Indeed, 
\citet{Gladstone09} have concluded based 
on the X-ray observations that
the strong outflow 
envelopes the inner part of the supercritical disk,
producing the cool spectral component 
in some ULXs.
Optically thick ($\tau_{\rm es} \gsim 3$)
low temperature ($\sim 10$ keV) corona 
are observed both in GRS 1915+105 
and ULX, Holmberg IX X-1
\citep{Kiki10a, Kiki10b}.

The geometrically thin, cold disk in model B is surrounded 
by the hot and rarefied matter (Figure \ref{fig2DB}).
Thus, the hot electrons in the corona would
Compton upscatter the seed photons from the cold disk.
While the thermal component of the spectra
in active galactic nuclei (AGNs) 
and BHBs is thought to be of disk origin,
the Compton upscattering in the corona
can account for the non-thermal hard component
\citep{Deufel00,Liu03,Done04,Done05}.
No such extended disk corona was reproduced 
by the simulations of the local patch of the disk,
which only show small, hot spots (with $T_{\rm gas}\sim 10^8$ K)
appearing near the disk surface \citep{Hirose06}.

In model B simulation, the optically thick, geometrically thin disk 
is truncated around $r\sim 7R_{\rm S}$ (see Section \ref{overB}).
Such a disk truncation has been reported by recent observations 
\citep{Kubota04,Tomsick09,Yamada09}
and by theoretical spectral modeling
\citep{Kawabata10}.
The blackbody radiation
cannot be emitted within the truncation radius,
but instead the seed photons generated in the outer cool disk 
will be Compton upscattered by the hot electrons
in the inner region.
It then follows that the iron line is produced 
not at the innermost stable circular orbit
but at the inner edge of the cool disk (truncation radius).
If so, it would be difficult to derive the information 
about the black hole spin \citep{Done06}.

Note, however, that evaporation of the disk gas due to the
thermal conduction is not taken into account
in the present simulation,
although it could be an important ingredient
to cause disk truncation
\citep[see Section 9.2.3 of][for comprehensive discussion]{KFM08}.

The disk in model C is optically thin and is composed of 
hot rarefied plasma.
Synchrotron self-absorption, synchrotron self-Compton,
and free$-$free emission should be dominant
radiative processes
\citep{Liu03,O05a,Kato09}.
Such spectra are observed in low-luminosity AGNs and
in low$-$hard state of BHBs. 
Sgr A* is also a good example.
Note that
the electron temperature has been reported to deviate from
the ion temperature 
\citep{NY05,Nakamura96,Manmoto97},
although we assume one-temperature plasma in the present study.
The emission of nonthermal electrons is thought to be non-trivial
\citep{Yuan03}.
More detailed study is, however, beyond the scope of this paper.

\subsection{Future Work}
Global RMHD simulation is a relatively new research field,
and, so there are plenty of issues to be explored in future.
We stress again that 
higher resolution simulations are needed to
study the detailed structure of the geometrically thin disk like model B.
The three-dimensional simulations should also be explored in future work.
It is well known that no magnetic dynamo works
in the axisymmetric calculations. Hence,
three dimensional simulations are indispensable
to make progress.
We also stress that it is better to relax reflection symmetry 
with respect to the equatorial plane, since such symmetry 
prevents the flows from going across the equatorial plane. 
We plan such simulations as future work.
In addition, we should take into account the relativistic effects,
since the outflow velocity becomes 10\%$-$70\%
of the speed of light near the rotation axis.
In the present work, we start calculations with a rotating torus, 
in which the magnetic fields are purely poloidal.
We should investigate the inflow-outflow structure 
starting with other initial conditions in future.

\begin{figure}
 \epsscale{0.9}
 \plotone{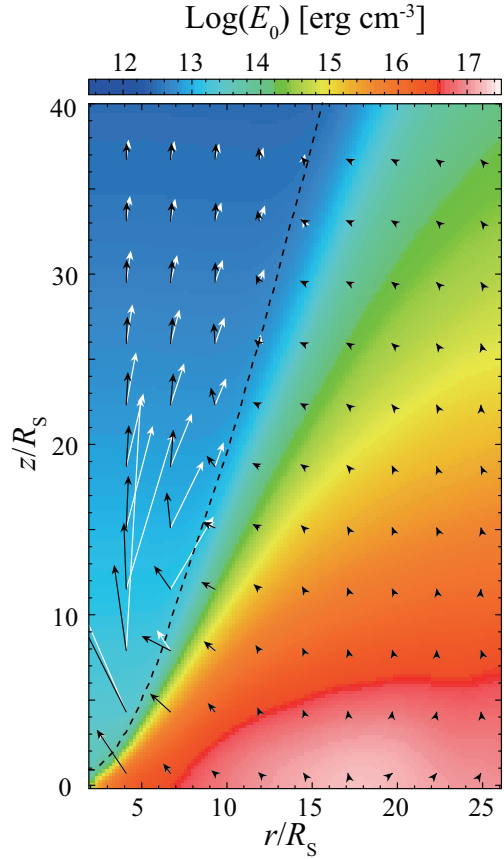}
 \caption{
 Radiative flux vectors by FLD approximation (black; FLD flux) and
 by integrating of radiation intensity (white; RT flux).
 Here, we solve the radiation transfer equation
 along $10^4$ light rays at each point for RT flux.
 Color contour indicates the radiation energy density,
 and the photosphere is plotted by dashed line.
 \label{RT_FLD}   
 }
\end{figure}
We used the FLD technique in the present study.
This is quite a useful technique and gives appropriate
radiation fields within the optically thick disk,
but we should keep in mind
that it does not always give precise radiation fields 
near the disk surface and above the disk
(where the optical depth is around unity or less).
In order to investigate the validity of the FLD,
we calculate accurate radiative flux (RT flux)
by solving radiation transfer equation 
along $10^4$ light rays at some points.
The resulting RT flux is presented by a white vector
in Figure \ref{RT_FLD}, where the radiative flux evaluated 
based on the FLD method (FLD flux) is overlayed by a black vector.
Here we employ time-averaged (6$-$7 s) structure of model A,
since the radiation force (radiative flux) in this model 
plays important roles.
We find in this figure that the FLD flux shifts from 
the RT flux in the outflow regions and near the photosphere, 
although the FLD flux is almost equal to RT flux 
in the optically thick disk regions.
Especially, the radial component of the FLD flux is 
negative (inward flux), but, in contrast, 
that of the RT flux is positive (outward flux,
e.g., see the fluxes at $r\sim 7R_{\rm S}$ with $z\sim 12Rs$).
The radiation from the vicinity of the black hole
works to produce the outward RT flux.
However, the inward flux appears by the FLD method,
since the direction of the flux is
determined by the gradient of the radiation energy density.
Thus, we should employ a more accurate method for the radiation fields
in future.
One of the improved methods is the so-called M1 closure scheme \citep{M1}.
In this method,
the 0th and 1st moment equations of radiation transfer are solved
to update the radiation energy and the radiative flux.
A closure relation,
which is used to evaluate the radiation pressure tensor,
is described in terms of the radiation energy and radiative flux
(In the FLD approximation, by contrast,
the radiation pressure tensor is prescribed by the radiation energy only,
and no information regarding radiative energy flux is used.
This may result in somewhat inaccurate evaluation regarding
the direction of radiation force in the jet acceleration region.)


Throughout the present study, 
we use the frequency-integrated energy equation of the radiation.
In order to obtain emergent spectra,
we should perform the frequency-dependent RMHD simulations.
However, such simulations are so heavy that 
it is practically impossible to perform.
Further,
we only took into consideration
the gas$-$radiation interaction via 
free$-$free and bound$-$free emission/absorption in our work.
However, synchrotron emission/absorption and Compton scattering
are important radiative processes.
They would play an important role
in the hot and tenuous regions 
found in accretion flows
as disks with lower accretion rate (model C) 
and outflows in all models.

\section{Conclusions}
With a two-dimensional global RMHD code, we could 
reproduce three distinct inflow-outflow modes 
around black holes by adjusting a density normalization.
Our three models with high, moderate, and low
density normalizations correspond to 
the two-dimensional RMHD version of the slim disk
(supercritical flow),
the standard disk, and the RIAF, 
all with substantial outflows.

We find the supercritical disk accretion flow,
of which the photon luminosity exceeds the Eddington luminosity,
in the case of the high density normalization (model A).
The vertical component of the radiation force 
balances with that of the gravity in the disk region of $z/r\lsim 2$
but it largely exceeds the gravity above the disk, $z/r\gsim 2$.
Our RMHD simulations reveals a new type of jet; 
i.e., the radiatively driven, magnetically collimated outflow,
which might account for the jets of radio-loud NLS1s and microquasars.
The disk, of which temperature is around $10^{7-8}$ K,
is surrounded by the hot outflowing matter, $> 10^9$ K,
which would induce
the Compton upscattering and 
the obscuration of the inner part of the disk.
Because of the mildly collimated radiative flux,
the apparent (isotropic) photon luminosity is $\sim 22L_{\rm E}$, 
which is $1.5\times 10^{41}\, \rm erg\, s^{-1}$
for the black hole of $50M_\odot$,
in the face-on view.
Even higher isotropic luminosity is feasible,
if a greater amount of material is supplied to the black hole.
Our supercritical model will be able to resolve 
the issue of the central engine of ULXs.

When the moderate density normalization is employed,
the radiative cooling is so effective that
the cold geometrically thin disk forms.
This cold disk with $\sim 10^6$ K 
is truncated at around $7R_{\rm S}$
and enveloped by the hot rarefied atmosphere with $T_{\rm gas}>10^9$ K,
Compton upscattering the seed photons from the cold disk.
The cold thermal component and the non-thermal hard component
of the spectra are observed in luminous AGNs and 
in the high-soft state of BHBs.
The disk wind appears above and below the disk, 
which was not predicted in the framework of the standard disk model.
The magnetic pressure in the vertical direction is responsible
for launching the gas from the disk surface.
This result is consistent with the observations of 
the blueshifted absorption lines.

The simulations with low density normalization
corresponds to the RIAF.
The magnetic-pressure force in cooperation with the gas-pressure force
drives the outflows.
The flow releases the energy via jets rather than via radiation.
The accretion flow as well as the outflows are hot and optically thin.
Thus, the spectra would resemble those
of the low-luminosity AGNs and of the BHBs
their low-hard state.

Finally, our simulations show that the vertically averaged disk
viscosity is roughly proportional to the total pressure 
\citep[see][]{O09}. 
However, the local energy dissipation rate is not simply
proportional to the gas (and radiation) pressures 
in the vertical direction at least in the models with 
high and low density normalization. 
The vertical profile of the dissipation rate is similar 
to that of the magnetic energy. 
This implies that the traditional
viscosity model is not perfectly correct.


\acknowledgments
We thank the anonymous reviewer 
for many helpful suggestions, which greatly improved
the original manuscript. 
The computations were performed on 
XT4 system at the Center for Computational Astrophysics (CfCA), 
National Astronomical Observatory of Japan (NAOJ).
This work is supported in part 
by the Ministry of Education, Culture, Sports, Science, and 
Technology (MEXT) Young Scientist (B) 20740115 (K.O.), 
by the Grant-in-Aid of MEXT (19340044, S.M.), 
by the Grant-in-Aid for the global COE programs on 
``The Next Generation of Physics, Spun from Diversity and Emergence''
from MEXT (S.M.).


\end{document}